\documentclass[twocolumn]{aastex631}
\usepackage{amsmath}
\usepackage{bm}

\newcommand{\vv}[1]{\bm{#1}}
\newcommand{\logalpha}[1]{$\alpha = 10^{-#1}$}

\usepackage[T1]{fontenc}

\graphicspath{{./}{figures/}}

\begin{document}

\title{Three-dimensional transport of solids in a protoplanetary disk containing a growing giant planet}

\author[0000-0002-5954-6302]{Eric Van Clepper}
\email{ericvc@uchicago.edu}
\affiliation{Department of the Geophysical Sciences, University of Chicago, Chicago, IL 60637, USA}

\author[0000-0002-3286-3543]{Ellen M. Price}
\altaffiliation{Heising-Simons Foundation 51 Pegasi b Postdoctoral Fellow}
\affiliation{Department of the Geophysical Sciences, University of Chicago, Chicago, IL 60637, USA}

\author[0000-0002-0093-065X]{Fred J. Ciesla}
\affiliation{Department of the Geophysical Sciences, University of Chicago, Chicago, IL 60637, USA}

\begin{abstract}
\noindent
We present the results of combined hydrodynamic and particle tracking post-processing modeling to study the transport of small dust in a protoplanetary disk containing an embedded embryo in 3D.
We use a suite of FARGO3D hydrodynamic simulations of disks containing a planetary embryo varying in mass up to 300 M$_\oplus$ on a fixed orbit in both high and low viscosity disks. We then simulate solid particles through the disk as a post-processing step using a Monte Carlo integration, allowing us to track the trajectories of individual particles as they travel throughout the disk.
We find that gas advection onto the planet can carry small, well-coupled solids across the gap opened in the disk by the embedded planet for planetary masses above the pebble isolation mass. This mixing between the inner and outer disk can occur in both directions, with solids in the inner disk mixing to the outer disk as well.
Additionally, in low viscosity disks, multiple dust pile-ups in the outer disk may preserve isotopic heterogeneities, possibly providing an outermost tertiary isotopic reservoir.
Throughout Jupiter’s growth, the extent of mixing between isotopic reservoirs varied depending on dust size, gas turbulence, and the Jovian embryo mass.
\end{abstract}

\keywords{Hydrodynamical simulations (767), Meteorites (1038), Protoplanetary disks (1300), Solar system astronomy (1529)}

\section{Introduction} \label{sec:intro}

As giant planets form, they sculpt their surroundings through gravitational interactions, creating rings, gaps, and spiral structures throughout the protoplanetary disks (PPDs) in which they are embedded. Such rings and gaps have been observed in numerous extrasolar PPDs \citep[e.g.][]{andrews_disk_2018, oberg_molecules_2021}, and in some cases the observed gas dynamics have been used to infer the existence of exoplanets in these disks \citep{pinte_kinematic_2018, pinte_kinematic_2019, teague_kinematical_2018, izquierdo_new_2022}. The flows of gas that develop as a result of these interactions play important roles in the dynamics of small dust and pebbles in the disk, as their trajectories are closely tied to the gas structure. Modeling these interactions in detail is key to understanding the formation history of planetesimals and planets in (extra-)solar systems.

The role that these substructures play in the compositional diversity of planetary materials is uncertain. In our own solar system, measurements of multiple isotopic systems have revealed a clear dichotomy between the carbonaceous (CC) and non-carbonaceous (NC) meteorites, including both chondritic and nonchondritic meteorites \citep{warren_stable-isotopic_2011, budde_molybdenum_2016, kruijer_age_2017}.
This dichotomy has been hypothesized to be the result of the preservation of two distinct compositional reservoirs in the solar nebula via a gap opened by a growing Jovian embryo \citep{kruijer_age_2017, gerber_mixing_2017, ebert_ti_2018, kruijer_great_2020}. Such a gap can create a pressure ``bump'' (a local pressure maximum) exterior to the planet where the gas pressure gradient is zero and the orbital velocity is Keplerian. In this region, solids and the surrounding gas co-orbit at the same speed, and the drag induced inward drift of solids halts. This can result in the creation of ring-like structures exterior to the planet at the local pressure maximum as solids drift into this region from beyond, depleting the exterior regions of the disk as solids pile up at the bump \citep[see e.g.][]{pinilla_ring_2012, dullemond_disk_2018}. 
In the solar system, this ring and gap substructure may be responsible for halting the inward drift of CC-like material and calcium-aluminum-rich inclusions (CAIs) from the outer disk to the inner disk, preserving the isotopic dichotomy observed in the meteorite record \citep{lambrechts_separating_2014,kruijer_age_2017,desch_effect_2018, kruijer_great_2020}. 

In addition to the bulk isotopic differences measured in meteorites, isotopic differences have also been measured in individual refractory inclusions within meteorites. For example, CAIs and sodium-aluminum-rich chondrules, two types of refractory inclusions found in both ordinary and carbonaceous chondrites, show distinct Ti isotopic ratios from one another, further indicating two reservoirs in the disk from which these refractory inclusions may have formed \citep{gerber_mixing_2017, ebert_ti_2018}. In addition to these isotopic differences, NC and CC meteorites have different distributions of CAIs, with CC chondrites tending to have both larger and a higher volume percent than NC chondrites \citep{dunham_calciumaluminum-rich_2023}. 

Although Jupiter is often invoked as the separator of these two reservoirs, some studies have brought into question the extent to which the embedded Jovian embryo could have prevented the exchange of material between them. In 1D models, while large solids remain trapped at pressure bumps, accretion flows onto the star through the disk may drag sufficiently small particles across pressure bumps, mixing outer and inner reservoirs \citep{chambers_rapid_2021}. In 2D, the interactions between the planet and disk result in non-axisymmetric substructures in the disk that further affect the trajectories of small solids in the disk. For example, interactions between the embedded giant planet and surrounding disk can create more complex substructures than a simple ring and gap, such as spiral wakes and vortices \citep{zhu_particle_2014, bae_formation_2017, bae_planet-driven_2018}. The asymmetry in the disk near the planet can have important implications for dust dynamics, in particular the transport of smaller solids from the outer reservoir to the inner nebula, also referred to as the ``filtering'' of material across the planet \citep{weber_characterizing_2018, haugbolle_probing_2019, stammler_leaky_2023}. 
Such filtering also depends on the gas advection through the disk and the planet's migration through the disk \citep{morbidelli_situ_2023}.
In 3D, the planet creates meridional flows, driving gas away from the planet at the midplane and onto the planet from the upper regions of the disk, such flows have been observed in real protoplanetary disks, indicating a promising method for identifying new embedded planet candidates \citep{morbidelli_meridional_2014, szulagyi_accretion_2014, fung_gap_2016, teague_meridional_2019, szulagyi_meridional_2022}. The consequences of this 3D structure created by an embedded planet on the subsequent transport of solids in the disk remains uncertain.

To date, models investigating the dynamics of solids in PPDs generally treat both the gas and dust as fluids, solving the continuity equations on a grid, giving gas and dust densities and velocities at fixed locations throughout the disk \citep{szulagyi_accretion_2014, krapp_3d_2022, szulagyi_meridional_2022, binkert_three-dimensional_2023}.  While these approaches are useful for determining large scale gas and dust structures through the disk, it is difficult to determine the detailed histories --- that is, the physical conditions experienced --- of individual particles through the disk, and the pathways through which material may or may not cross the gap opened by an embedded giant planet.

Here, we use a particle tracking technique to study the 3D dynamics of small solids in a disk containing an embedded planetary embryo.
This approach allows us to track the histories of small dust and pebbles in a protoplanetary disk during different stages of a Jupiter-like planet's growth. Specifically, this approach allows us to connect different regions of the disk, identifying not only regions of the disk where solids concentrate, but also from where those solids may have originated in the disk. We can also identify the mechanisms by which potential mixing between reservoirs in the disk occurs and the extent of mixing based on disk and grain parameters.

While we focus specifically on an embedded Jovian embryo and implications for planetesimals in the solar nebula, the results presented here are applicable to protoplanetary disks broadly. In Section \ref{sec:method} we describe the hydrodynamic simulations of the gaseous disk containing the embedded planet and the Monte Carlo integration technique used to track the solid particles. In Section \ref{sec:results} we present the results of this suite of models and discuss their implications within a Solar System context in Section \ref{sec:discussion}. In Section \ref{sec:conclusion} we summarize our results and present avenues for future research.

\section{Methodology} \label{sec:method}

In this work, we combine FARGO3D hydrodynamic simulations \citep{masset_fargo_2000, benitez-llambay_fargo3d_2016} with the particle tracking technique of \citet{ciesla_residence_2010, ciesla_residence_2011} to track the evolution of small dust through a PPD sculpted by an embedded planetary embryo. We first simulate the gas disk as a single fluid in FARGO3D until a steady state is reached. We explore a variety of disk conditions including a range of planet masses and disk viscosities.
We then simulate a population of dust particles in a post-processing step, holding the gas density and velocity field constant over the particle integration. Particle sizes studied here range from 10 \textmu m to 1 cm, representing the approximate range of dust expected to be found in the protosolar nebula, including CAIs and other meteorite constituents \citep[see e.g.][]{dunham_calciumaluminum-rich_2023}. Thus, we expect these results to be informative of the dynamics of the range of dust sizes in the solar nebula. A full list of all the different models considered is given in Table \ref{tab:all_models}.

\begin{deluxetable}{ccc}
\label{tab:all_models}
\tablecaption{Simulations}
\tablehead{
    \colhead{Planet Mass, $M_{pl}$} & \colhead{Viscosity} & \colhead{Particle Sizes, $s$} \\
    \colhead{$M_\oplus$} & \colhead{$\log_{10} \alpha$} & \colhead{cm}
}
\startdata
0 & -3, -4 & 1, 0.1, 0.01, 0.001 \\
1 & -3 & 1, 0.1, 0.01, 0.001 \\
5 & -3 & 1, 0.1, 0.01, 0.001 \\
10 & -3, -4 & 1, 0.1, 0.01, 0.001 \\
20 & -3, -4 & 1, 0.1, 0.01, 0.001 \\
50 & -3, -4 & 1, 0.1, 0.01, 0.001 \\
100 & -3, -4 & 1, 0.1, 0.01, 0.001 \\
200 & -3 & 1, 0.1, 0.01, 0.001 \\
300 & -3, -4 & 1, 0.1, 0.01, 0.001 \\
\enddata
    
\end{deluxetable}

\subsection{Hydrodynamic Simulations}\label{sec:method-hd}

Using the publicly available version of FARGO3D \citep{masset_fargo_2000, benitez-llambay_fargo3d_2016}, we use a 3D, spherical mesh to simulate a gaseous disk around a solar-mass star.  The mesh has $2048 \times 256 \times 32\ (N_\phi \times N_r \times N_\theta)$ cells, with logarithmic spacing in the radial direction, and the mesh being defined on $\phi \in (-\pi,\pi)$, $r/R_0 \in (0.5,3.0)$, and $\theta \in (\pi/2-0.15, \pi/2)$. Within each mesh, we include an embedded planet with constant mass, $M_{pl}$, varying from 0 to 300 $M_\oplus$. In all cases, we place the planet at the midplane, $\theta=\pi/2$, and at the azimuthal angle $\phi=0$. We define the physical radius of the planet's orbit to be $R_{pl} = R_0 = 5.2\mathrm{\ au}$. 
The planet's orbit is held fixed at this location throughout the hydrodynamic simulation and particle integration.
We use a rotating reference frame such that the planet remains at $\phi=0$ throughout the simulation.

The initial surface density of the disk at $r = R_0$ is 
$\Sigma_0 \approx 210 \mathrm{\ g\ cm}^{-2}$, and it scales radially as 
\begin{equation}
    \Sigma(r) = \Sigma_0\left(\frac{r}{R_0}\right)^{-1}.
\end{equation}
We set the temperature in the disk to be $T(R_0)=T_0\approx118 K$, and it scales radially as $T \propto r^{-1/2}$. These values correspond to a surface density and temperature at 1 au of $\Sigma(1 \text{ au}) = 1088$ g cm$^{-2}$ and $T(1 \text{ au}) = 270\ \text{K}$ respectively. The scale height of the disk at $r=R_0$ is $H(R_0) = H_0 = 0.05R_0$ with a flaring index of $1/4$; that is, the gas scale height is given by
\begin{equation}
    H(r) = \frac{c_s}{\Omega_k} = 0.05r\left(\frac{r}{R_0}\right)^{1/4}
\end{equation}
where $c_s$ is the sound speed of the gas and $\Omega_K = \sqrt{GM/r^3}$ is the Keplerian orbital frequency. As such, the FARGO3D mesh extends approximately 3 scale heights above the midplane. The kinematic viscosity of the disk is then given by $\nu = \alpha c_s H$ as in \citet{shakura_black_1973}, where we use $\alpha = 10^{-3}$ and $10^{-4}$ in our models.
The temperature profile is held constant throughout the simulation and the disk is vertically isothermal. 

The disk is initially axisymmetric and in hydrostatic equilibrium, with the equation for density throughout the disk given in the appendix \ref{sec:3d-structure}. In the co-rotating frame with the planet, the rotation rate of the frame, $\Omega_f$, is equal to the orbital frequency of the planet, $\Omega_0$.

Each model is integrated for at least 500 orbits of the planet ($\sim 5900$ years), or until a steady state in gas density is reached.  We define a steady state in the disk as being reached when the azimuthally averaged midplane density everywhere in the disk changes by less than 1\% over the course of one orbit.
For some models a steady state is not reached before 1000 orbits, and the midplane experiences small oscillations over orbital timescales. For these cases, we use a time average of the ten orbits from 990 to 1000 for the particle integration, following \citet{fung_gap_2016}. Although this approach may ignore some transient structures such as vortices at the gap edges, the timescale of these orbit-to-orbit variations are short compared to the integration time of the particles, and vortex lifetimes for planets smaller than $1 M_\text{Jup}$ tend to be shorter than the timescales considered here \citep{hammer_slowly-growing_2017}. The effect of this time-varying density on particle transport can be explored in future work.

\subsection{Particle Integration}\label{sec:method-part}

Particles are taken to be compact spheres with a given constant size, $s$, and solid density, $\rho_s = 2$ g cm$^{-3}$. Each particle's initial radial and azimuthal velocity is determined from analytic particle velocities for an axisymmetric disk based on the surrounding gas velocity \citep{takeuchi_radial_2002, armitage_lecture_2017}.

The particle is forward integrated following the technique of \citet{ciesla_residence_2010, ciesla_residence_2011} extended to 3D, non-axisymmetric structure. The particle tracking is done in Cartesian coordinates, with the gas density and velocity at the position of the particle interpolated from the FARGO3D outputs.
Using this particle tracking technique, as opposed to fluid or dye-tracking approaches, allows us to follow the trajectories of individual particles, seeing how solids may transit across different regions of the disk. In the current public release of FARGO3D, solids are treated as pressureless fluid of constant Stokes number. Here, we use a constant particle size in the integration, thus the drag on a given particle can be consistently determined at each time and location, adjusting the particle's Stokes number as it moves through different environments. As a particle moves through the disk, the surrounding gas density, and thus particle Stokes number, can vary by multiple orders of magnitude, and accounting for this difference can have important consequences for resulting solid dynamics \citep[][Price et al. \textit{in press}]{weber_characterizing_2018}.

In our model, each particle's location and velocity is integrated forward in time using an Eulerian approach:

\begin{equation}\label{eq:main_solve}
    \vv{Y}^i = \vv{Y}^{i-1} + \frac{\partial}{\partial t}\vv{Y}^{i-1}\Delta t + \vv{R}\left[6 D(\vv{x}') \Delta t\right]^{1/2}
\end{equation}
where 
\begin{equation}
    \vv{Y}^i = \begin{bmatrix}
        x^i \\ y^i \\ z^i \\
        v_x^i \\ v_y^i \\ v_z^i
    \end{bmatrix}
\end{equation}
is a 6D Cartesian position and velocity vector. We can also write $\vv{Y}^i = [\vv{x}^i, \vv{v}^i]$ where $\vv{x}^i = [x^i,y^i,z^i]$ and $\vv{v}^i = [v_x^i, v_y^i, v_z^i]$, giving the Cartesian position and velocity (respectively) of the particle in the corotating frame centered at the center of mass of the system. The superscript $i$, denotes the values of $\vv{Y}$ at the $i$-th timestep. Where it is not necessary, we drop the $i$ superscript for simplicity. The last term in equation \eqref{eq:main_solve} represents a random displacement over timestep $\Delta t$ associated with diffusion, with $\vv{R} = [p_x,p_y,p_z,0,0,0]$ where $p_{[x,y,z]}$ is a random number drawn from a uniform distribution ranging from [-1,1].
$\vv{x}'$ is a location chosen near $\vv{x}^{i-1}$ offset toward increasing diffusivity and is given by

\begin{equation}
    \vv{x}' = \vv{x}^{i-1} + \frac{1}{2}\vv{\nabla}D(\vv{x}^{i-1})\Delta t
\end{equation}
and $D(\vv{x})$ is the diffusivity of the particle at a given location \citep[see discussion in][]{ciesla_residence_2010}. Thus, we assume diffusion leads to a random displacement of the particle without changing its velocity. The particle diffusivity is related to the gas diffusivity using the relationship given in \citet{youdin_particle_2007}:
\begin{equation}\label{eq:part_diff}
    D = \frac{D_g}{1+\text{St}^2}.
\end{equation}
where St $= t_\text{stop}\Omega_K$ is the Stokes number and $t_\text{stop} = \rho_s s/\rho_g c_s$ is the stopping time of the particle.
The gas diffusivity is the same as the kinematic viscosity $D_g = \nu = \alpha c_s H$, e.g. we assume a Schmidt number of unity.

The time derivative of $\vv{Y}$ is

\begin{equation}
    \frac{\partial}{\partial t}\vv{Y} = \begin{bmatrix}
        v_{\mathrm{eff},x} \\
        v_{\mathrm{eff},y} \\
        v_{\mathrm{eff},z} \\
        a_x \\
        a_y \\ 
        a_z
    \end{bmatrix} = 
    \begin{bmatrix}
        \vv{v}_\mathrm{eff} \\
        \vv{a}
    \end{bmatrix}.
\end{equation}
The effective velocity, $\vv{v}_\mathrm{eff} = [v_{\mathrm{eff},x}, v_{\mathrm{eff},y}, v_{\mathrm{eff},z}]$ accounts for the actual velocity of the particle plus effects due to the spatially varying density and diffusivity throughout the disk \citep[see][]{ciesla_residence_2010, ciesla_residence_2011},

\begin{equation} \label{eq:veff}
    \vv{v}_\mathrm{eff} = \vv{v} + \frac{D(\vv{x})}{\rho_g(\vv{x})}\vv{\nabla}\rho_g(\vv{x}) + \vv{\nabla}D(\vv{x}).
\end{equation}

The acceleration, $\vv{a} = [a_x,a_y,a_z]$, is determined from the gravity due to the planet, gravity from the central star, drag with the surrounding gas, and the centripetal acceleration in the rotating frame. 
That is, the total acceleration of the particle is

\begin{equation}
    \vv{a} = \vv{a}_\mathrm{grav} + \vv{a}_\mathrm{grav,pl} + \vv{a}_\mathrm{drag} + \vv{a}_\mathrm{rot}.
\end{equation}
Tracer particles do not interact with one another, nor do we account for back reactions of the dust on the gas.

The gravitational acceleration due to the central star is given by:
\begin{equation}
    \vv{a}_\mathrm{grav} = -\frac{GM_\star}{|\vv{x}-\vv{x}_\star|^3}(\vv{x}-\vv{x}_\star),
\end{equation}
and the gravitational acceleration from the planet is:
\begin{equation}
    \vv{a}_{\mathrm{grav},\text{pl}} = -\frac{GM_\text{pl}}{|\vv{x}-\vv{x}_\text{pl}|^3}(\vv{x}-\vv{x}_\text{pl}),
\end{equation}
where $M_\star$ and $\vv{x}_\star$ are the mass and position of the central star, and $M_\text{pl}$ and $\vv{x}_\text{pl}$ are the mass and position of the planet respectively.

The drag acceleration of the particle is determined following the methodology of \citet{tanigawa_accretion_2014}:
\begin{equation} \label{eq:full_drag}
    \vv{a}_\mathrm{drag} = \frac{\vv{F}_\mathrm{drag}}{m} = -\frac{(C_D/2)\pi s^2 \rho_g |\vv{u}|\vv{u}}{(4/3)\pi s^3 \rho_s},
\end{equation}
where $C_D$ is the drag coefficient, $\rho_g$ is the local gas density, $\vv{u} = \vv{v} - \vv{v}_\mathrm{gas}$ is the relative velocity of the particle, and $m$ is the mass of the particle.

Equation \ref{eq:full_drag} simplifies to the following form:
\begin{equation}
    \vv{a}_\mathrm{drag} = -\frac{3}{8}\frac{C_D}{s}\frac{\rho_g}{\rho_s}|\vv{u}|\vv{u}.
\end{equation}
In the Epstein drag regime, which we use for all cases as the particle sizes are less than the mean free path in the gas, this is further simplified to:

\begin{equation}\label{eq:epdrag}
    \vv{a}_\mathrm{drag} = -\frac{\rho_g c_s}{\rho_s s}\vv{u} = 
    -\frac{\vv{u}}{t_\text{stop}}.
\end{equation}

Finally, the acceleration due to rotation, including the Coriolis and centrifugal forces is given by
\begin{equation}
    \vv{a}_\mathrm{rot} = -2\vv{\Omega}_f \times \vv{v} - \vv{\Omega}_f \times (\vv{\Omega}_f \times \vv{x})
\end{equation}
or, in expanded vector notation
\begin{equation}
    \vv{a}_\mathrm{rot} = 
    \begin{bmatrix}
        2v_y\Omega_f + x\Omega_f^2 \\
        -2v_x\Omega_f + y\Omega_f^2 \\
        0
    \end{bmatrix}
\end{equation}
where the $\vv{\Omega}_f = [0,0,\Omega_f]$ is the rotation rate of the frame.

While particles are integrated forward in time, their positions, velocities, and Stokes numbers are recorded at each timestep. If particles leave the bounds of the mesh (i.e. drift to the inner edge or are lofted such that their polar angle exceeds $\pi/2\pm0.15$), they are removed from the simulation. Additionally, any particles that pass within the planetary envelope are considered accreted by the planet, and removed from the simulation. Following \citet{chambers_steamworlds_2017} and \citet{barnett_thermal_2022}, the planetary envelope radius is given by

\begin{equation}\label{eq:r_env}
    r_\textrm{env} = \min\left[\frac{r_\textrm{Hill}}{4}, r_\textrm{Bondi}\right]
\end{equation}
where $
r_\textrm{Hill} = a_{pl}(M_{pl}/3M)^{1/3}
$
is the Hill radius of the planet and
$
r_\textrm{Bondi} = 2GM_{pl}/c_s^2
$
is the Bondi radius of the planet.

We use a Runge-Kutta 4\textsuperscript{th} order method for integration, with the timestep, $\Delta t$, chosen to ensure numerical stability. The timestep is defined as
\begin{equation}\label{eq:dtcalc}
    \Delta t = \mathrm{min}(0.1~ \mathrm{yr}, \Delta t_\mathrm{orbit}, \Delta t_\mathrm{drag}, \Delta t_\mathrm{diff}, \Delta t_\mathrm{rho})
\end{equation}
where

\begin{equation}
    \Delta t_\mathrm{orbit} = \eta \frac{2\pi}{\Omega_K},
\end{equation}
\begin{equation}
    \Delta t_\mathrm{diff} = \xi \frac{|\vv{x}|}{\vv{|\nabla}D|},
\end{equation}
\begin{equation}
    \Delta t_\mathrm{rho} = \zeta \frac{\rho_g}{D}\frac{|\vv{x}|}{|\vv{\nabla}\rho_g|}.
\end{equation}

Here, each expression represents a step-size limit to ensure numerical stability. In $\Delta t_\mathrm{orbit}$, we limit the timestep to be a sufficiently small fraction, $\eta=1/50$, of the orbit, to account for changes in $v_x$ and $v_y$ with location. Additionally, we limit the step size of the particle during the integration dependent on the components of the effective velocity, where $\Delta t_\mathrm{diff}$ and $\Delta t_\mathrm{rho}$ are based on the diffusive terms of the effective velocity in equation \eqref{eq:veff}. We also include an upper limit of 0.1 yr in all cases, to prevent unphysical displacement due to turbulence, even in smooth regions of the disk.
The coefficients $\xi$ and $\zeta$ in $\Delta t_\mathrm{diff}$ and $\Delta t_\mathrm{rho}$ are chosen to be $10^{-6}$ and $10^{-5}$ respectively to ensure these timesteps remain small relative to the stopping time of moderately coupled solids. 
As the effective velocity of the particle depends on $\vv{\nabla}\rho_g$ and $\vv{\nabla}D$, the timestep of the integration is decreased where the magnitude of these values is large.
This ensures particles only move small distances in a given timestep when in regions of sharp density or diffusivity gradients.
This was tested using values of $\xi$ and $\zeta$ an order of magnitude smaller and there was no significant difference to the integration results.

In testing, we found that particles with Stokes number less than about 10$^{-3}$ were always strongly coupled to the gas, that is, $\vv{u} \approx 0$. However, these particles tended to have long integration times as $t_\textrm{stop}$ was small, limiting $\Delta t$. To allow for reasonable integration times, we use an adjusted $\Delta t_\textrm{drag}$ in equation \eqref{eq:dtcalc}.
The adjusted $\Delta t_\textrm{drag}$ is given by the equation

\begin{equation}
    \Delta t_\textrm{drag} = \frac{t_\textrm{stop}}{\sigma(\textrm{log}_{10}\textrm{St})}= \frac{s\rho_s}{c_s\rho_g} \frac{1}{\sigma(\textrm{log}_{10}\textrm{St})},
\end{equation}
where $\sigma$ is a sigmoid function centered at $-3$

\begin{equation}
    \sigma(x) = \frac{1}{1+e^{-(x+3)}}.
\end{equation}

Thus, for particles with Stokes number greater than $10^{-3}$, $\Delta t_\textrm{drag} \approx t_\textrm{stop}$. Alternatively, for particles with Stokes number much less than $10^{-3}$, $\Delta t_\textrm{drag}$ is large and integration steps can be much longer as particles follow the flow of the gas. When a particle's Stokes number is less than $10^{-3}$, the drag force is assumed to be zero and the particle velocity is set to the same as the surrounding gas, that is, $\vv{v} = \vv{v}_\textrm{gas}$. We tested different limits of this coupling Stokes number, and found no significant difference in results when this value was lowered by an order of magnitude.

\section{Results}\label{sec:results}

\subsection{Hydrodynamic Simulations}

\begin{figure}
    \centering
    \includegraphics[width=\linewidth]{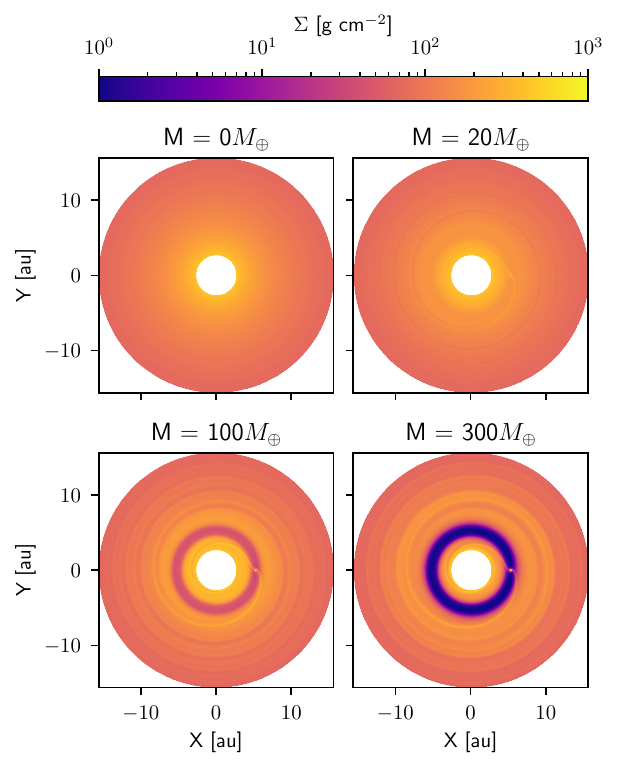}
    \caption{Surface density maps for FARGO3D simulations containing embedded planets of mass 0 (no planet), 20, 100, and 300 $M_\oplus$. Results are shown in the co-rotating frame with the planet, with the planet located at $x=5.2$ au and the central star at the origin. At 20 $M_\oplus$ a small gap can be seen, although the planet has not yet reached the pebble isolation mass. At 100 and 300 $M_\oplus$, the co-rotating spiral arms can be seen coming off the planet, with density perturbations from the initial surface density increasing with embryo mass.}
    \label{fig:surface_densities}
\end{figure}

\begin{figure}
    \centering
    \includegraphics[width=\linewidth]{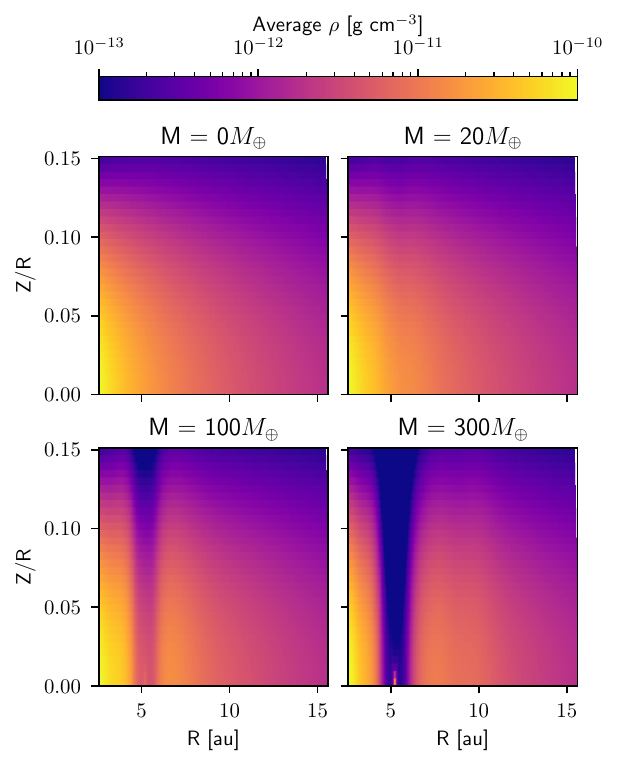}
    \caption{Azimuthally averaged densities for the same set of embedded planets as Figure \ref{fig:surface_densities}. The density is averaged over the full disk, $\phi \in [-\pi,\pi]$. The gap opened by the planet extends from the midplane ($Z/R = 0$) to the upper layers of the disk. The dense planetary embryo can be seen at the location of the planet.}
    \label{fig:side_densities}
\end{figure}

The results of the FARGO3D simulations for several embedded planet masses are shown in Figures \ref{fig:surface_densities} and \ref{fig:side_densities}. The results shown here and in the following sections are using the fiducial disk viscosity of $\alpha = 10^{-3}$; we discuss the effects of lowering disk viscosity in section \ref{sec:alpha4_results}.
The gap opened by the planet and spiral arm structure can be seen in the Figures.
As in previous studies \citep[e.g.][]{kanagawa_mass_2016, fung_gap_2016, duffell_empirically_2020}, the gap grows wider and deeper for more massive planets and with disk decreasing viscosity.

\begin{figure}
    \centering
    \includegraphics[width=\linewidth]{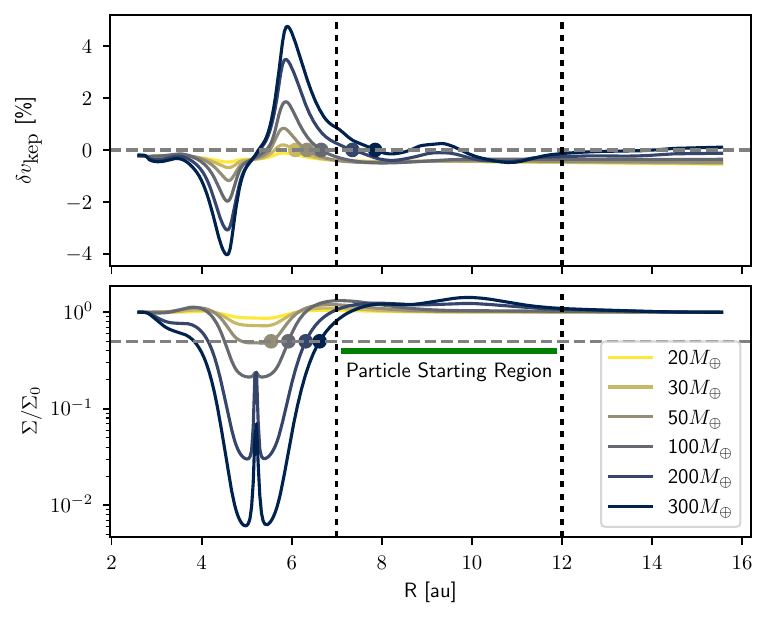}
    \caption{Azimuthally averaged midplane azimuthal velocity (\textit{top}) and surface density relative to a disk with no planet (\textit{bottom}). The azimuthal velocity is given as percent deviation from keplerian, $\delta v_{\textrm{kep}} = (v_\phi - v_\textrm{kep})/v_\textrm{kep}$, and the horizontal dashed line is at $\delta v_\textrm{kep}=0\%$. In the bottom panel, the gap edge is defined as where $\Sigma/\Sigma_0 = 0.5$. The region where we start particles in our simulation is shown between the two vertical dotted lines at 7 and 12 au, chosen to be outside the gap for all planet masses considered. }
    \label{fig:dvkep}
\end{figure}

Figure \ref{fig:dvkep} shows the azimuthally averaged deviation from Keplerian velocity and surface density depletion relative to the initial disk that develops as a result of the planet-disk interactions. In our model with $\alpha = 10^{-3}$, a planet mass $\gtrsim30 M_\oplus$ is necessary for gas velocities to reach Keplerian at the pressure maximum exterior to the planet, in good agreement with typical pebble isolation masses in the solar nebula found in other studies \citep{lambrechts_separating_2014, bitsch_pebble-isolation_2018}. Here, we define the gap as the region of the disk where the azimuthally averaged surface density is depleted by at least 50\%. While planetary masses greater than or equal to $30 M_\oplus$ are required to deplete the gas by more than 50\% to create a gap, we note that the gas is still depleted around the planet for smaller embryo masses, just not to a sufficient level to halt the inward drift of solids. The location of the gap edge is highlighted in Figure \ref{fig:dvkep}. In our simulations, we find that gap width, scales as $M_{pl}^{1/2}$ as in other studies \citep[e.g.][]{duffell_simple_2015, kanagawa_formation_2015, kanagawa_mass_2016, duffell_empirically_2020}.

\begin{figure}
    \centering
    \includegraphics[width=\linewidth]{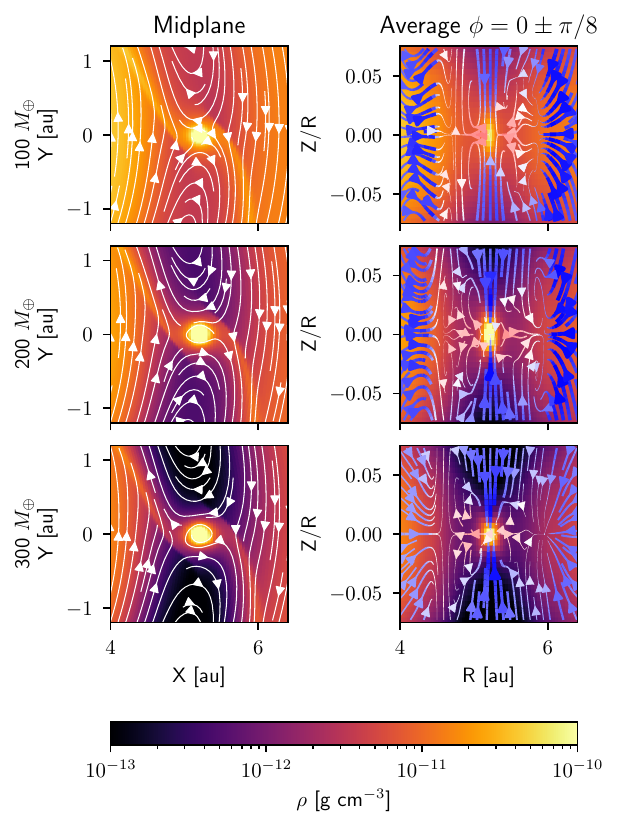}
    \caption{Results of FARGO3D simulation of disks containing a 100, 200, and 300 $M_\oplus$ (top to bottom) embedded planet showing the gas density and velocity.  The left panels show the Cartesian midplane gas structure of the disk. The right panels show the radial and vertical gas density and gas advection near the planet, averaged over $\phi = 0 \pm \pi/8$. Here, the arrow color corresponds to the advection of the gas relative to the planet, with blue indicating flows towards the planet and red away. As the planetary embryo grows, the magnitude of the gas advection velocities grow while the width and depth of the gap deepens.}
    \label{fig:fargo_results}
\end{figure}

The gas density and velocity around the embryo for three simulations with the largest masses, 100, 200, and 300 $M_\oplus$, are shown in Figure \ref{fig:fargo_results}. As the mass of the planet grows, the strength of the advective flows in the disk increase, with clear meridional flow circulation patterns developing around the embedded embryos.
This outward flow can be seen along the spiral arms in the left side of Figure \ref{fig:fargo_results}, and at the midplane near the planet in the right column.
The increased strength of these flows outward along the midplane result in both faster gas velocities and a greater radial extent of the transport of material away from the planet, in agreement with previous 3D studies of embedded disks \citep[e.g.][]{morbidelli_meridional_2014, fung_gap_2016, szulagyi_meridional_2022, krapp_3d_2022}.

\subsection{Particle Crossing From Outer Nebula to Inner Nebula}\label{sec:filtering}  

In the absence of any planet, the transport of mm- to cm-sized solid particles is largely dominated by inward drift due to gas drag \citep[e.g.][]{weidenschilling_aerodynamics_1977}. As the growing embryo opens a gap in the nebula, the inward drift of solids can be halted in the resulting pressure bump. However, as has been shown in previous 2D simulations \citep[e.g.][]{weber_characterizing_2018, haugbolle_probing_2019} small particles ($< 1$~mm) can filter past Jupiter-mass embedded embryos, replenishing the inner nebula with solids. Here, we consider particle movement in 3D to understand the details of these motions in the presence of an embedded planet. 

To explore the trajectories of these crossing particles and identify the size and provenance of such grains, we track the evolution of different sized particles that begin evenly distributed in the 7 to 12 au region through all disks considered here. We consider the outer disk to be the region exterior the orbit of the embryo at 5.2 au, and the inner disk the region interior to the orbit of the embryo.
We choose this initial location for the particles as it is just exterior to the gap opened by the 300 $M_\oplus$ embryo, the most massive planet and widest gap considered (see Figure \ref{fig:dvkep}).

\begin{figure*}
    \centering
    \includegraphics[width=\linewidth]{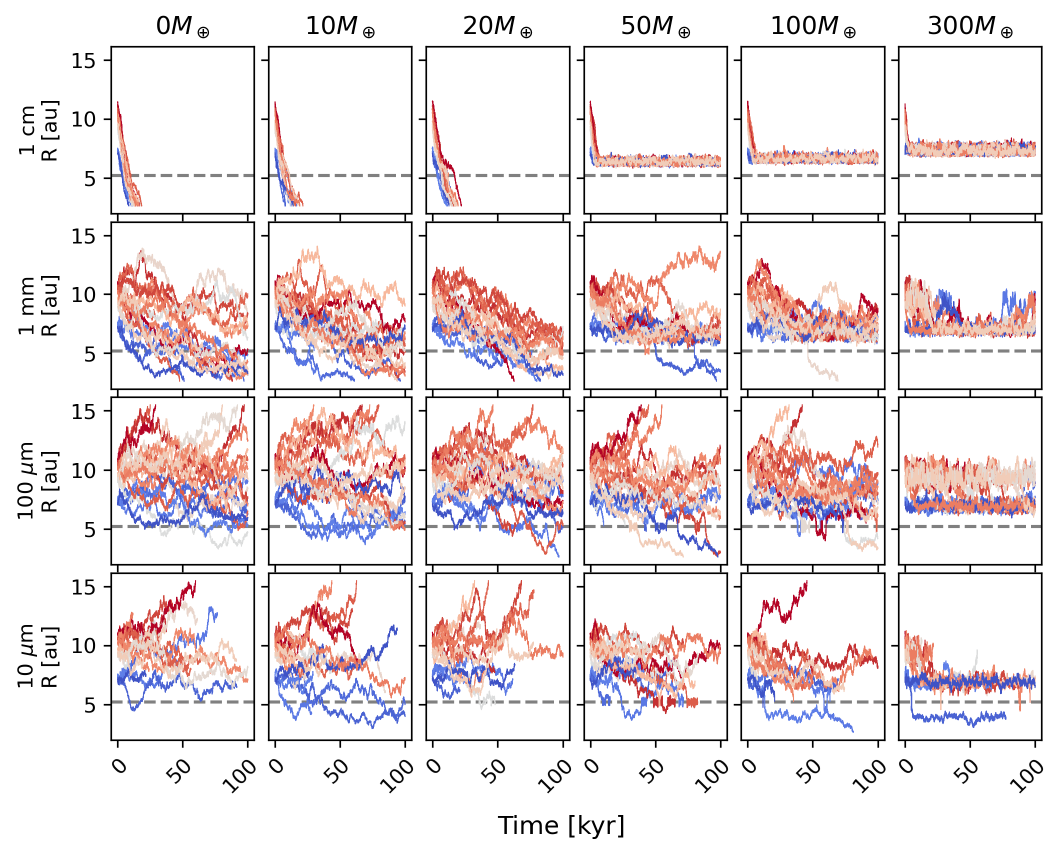}
    \caption{A selection of particle trajectories, showing the radial location of 20 particles as a function of time. Each subplot represents the evolution of particles in a disk containing different mass planetary embryos (columns) and different size grains (rows). Each line is colored according to the particle's initial location, ranging from 7 to 12 au, with bluer lines representing particles starting at 7 au and redder lines showing particles starting at 12 au. The location of the planet at 5.2 au is marked in each subplot by the horizontal dashed line. Broadly, large grains with in disks containing small embryo masses efficiently drift from the outer disk to the inner disk, while this inward drift is halted above the pebble isolation mass. Small particles may cross the planetary for large planetary embryo masses as a result of strong advection, either bringing the solids into the gap in a horseshoe orbit, to be accreted, or past the gap into the inner disk. In cases where particles do not survive for the full 100 kyr, they may be accreted onto the planet or diffuse outside of the domain of the simulation.}
    \label{fig:trajectories}
\end{figure*}

Sample trajectories of particles of different sizes for a variety of embryo masses are plotted in Figure \ref{fig:trajectories}, showing the radial movement of 20 illustrative particles over their 100 kyr integration time. As can be seen, the efficiency at which particles move from the outer disk to the inner disk depends on both grain size and embryo mass. Below the pebble isolation mass ($<30 M_\oplus$) the planetary embryo does not create an effective pressure bump outside of its orbit, allowing the 1 cm size particles to efficiently drift from the outer disk to inner disk. Smaller particles are well coupled with gas and have much longer drift timescales, and thus do not significantly drift inward over the 100 kyr simulated here; and instead their dynamics are dominated by diffusion.
Once the embryo is massive enough, inward drift is halted for all particles larger than 1 mm in size.

\begin{figure}
    \centering
    \includegraphics[width=\linewidth]{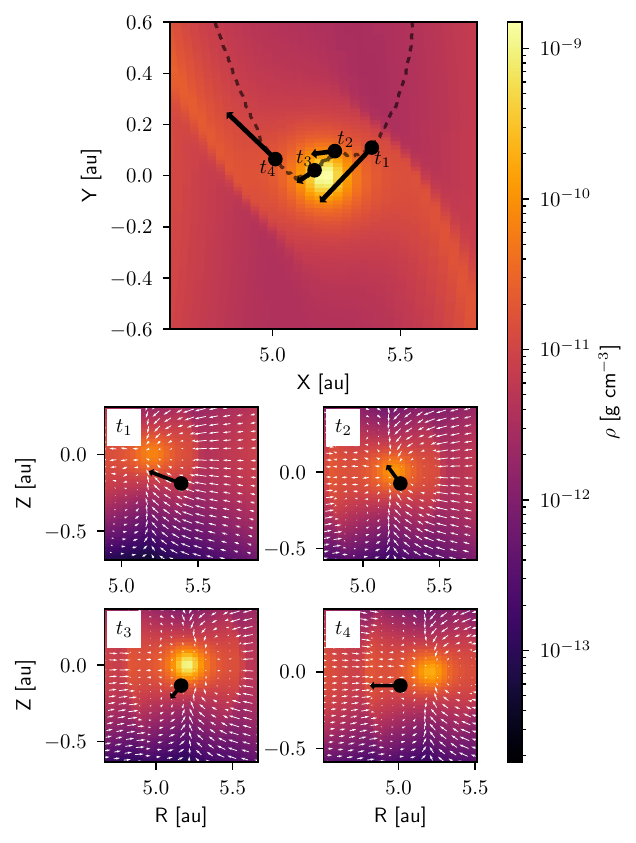}
    \caption{An example trajectory of a solid particle that crosses from the outer disk to the inner disk. The top panel shows the path followed by the particle as the black dashed line projected onto the midplane of the disk, with 4 locations and particle velocities labeled. The particle velocity and surrounding gas density and advection in the R-Z plane is shown below for each of the highlighted times in the top panel. For this particle, the inward advecting gas pulls the solid particle towards the planet ($t_1$ and $t_2$) following the meridional flow. The particle comes close to the planet, but remains outside of the accretion envelope ($t_3$). Then, the particle follows the outward meridional flow along the midplane to the inner disk ($t_4$).}
    \label{fig:onetrajectory}
\end{figure}

However, once the planetary embryo reaches $\gtrsim~100~M_\oplus$, the mass corresponding to the runaway accretion phase of giant planet formation \citep{helled_revelations_2022}, particles smaller than 100 \textmu m that are well coupled to these strong advective flows can pass the orbit of the embedded embryo. The path of a 10 \textmu m sized particle that moves past a 100 $M_\oplus$ embryo is shown in Figure \ref{fig:onetrajectory}. At the first time shown, the particle is exterior to the planet and approximately one scale height away from the midplane. Here, the gas advection is towards the planet, pushing the particle both inward and downward. The particle passes nearby the planet, $\sim 200 R_\text{Jup}$ or about $1.5r_\text{env}$, before moving to an orbit interior to the planet. Although the flow of gas at the midplane near the embryo is outward, if the solids diffuse to a high enough elevation, they can be entrained in the inward flowing gas. This favors crossing for small particles or disks with high $\alpha$.

\begin{figure}
    \centering
    \includegraphics[width=\linewidth]{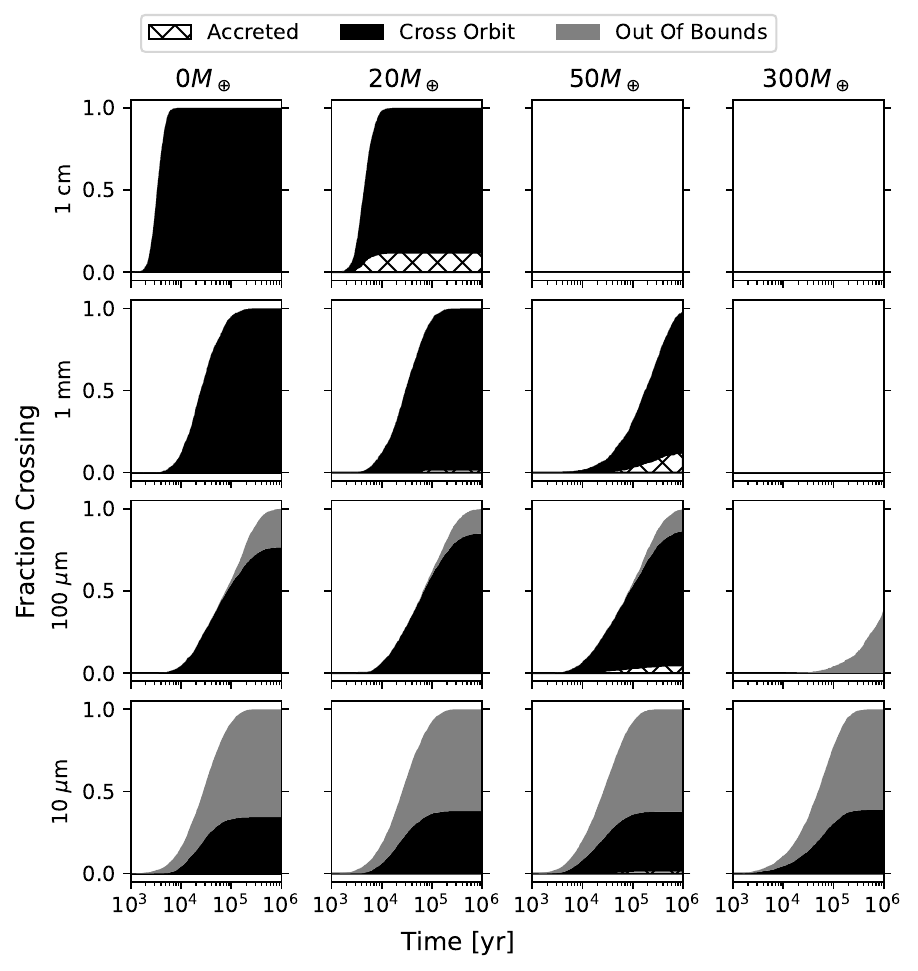}
    \caption{The fraction of solids starting at 7 au that cross the orbit of the planet. For the 1 cm and 1 mm sized pebbles, drift across the orbit of the planet occurs over longer timescales as the planet mass increases. Some grains are lost to accretion onto the planet, indicated by the cross hatched regions. Other grains diffuse outside of the domain of the integration, and are shown by the light gray region of the plots. For the smaller 100 and 10 \textmu m sized dust, which is well coupled to the gas, the fraction crossing increases with increasing planet mass. This is because the larger embryo creates advective flows through the gap, carrying the small dust with the gas. The final percentages for each planet mass and grain size are shown in Figure \ref{fig:pcfiltered}.}
    \label{fig:crosscdf}
\end{figure}

\begin{figure}
    \centering
    \includegraphics[width=\linewidth]{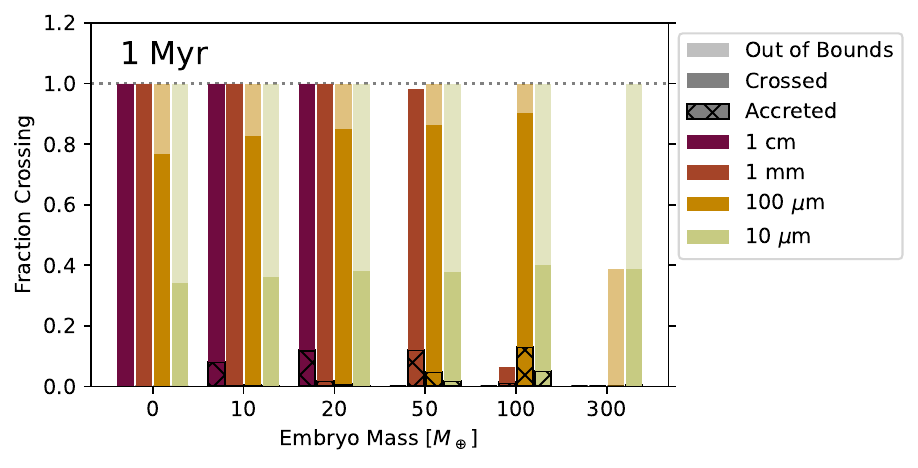}
    \caption{Of solids initialized at 7 au (the same population as shown in Figure \ref{fig:crosscdf}), we show the fraction of particles either accreted by the planet or crossing the orbit of the planet after 1 Myr of integration. While the planet is below the pebble isolation mass, the fraction of 1 cm particles crossing the orbit of the planet is near unity, with up to $\sim 10$\% accreted by the planet, as their dynamics are primarily driven by inward drift over the timescale of the integration. Smaller particles have much lower crossing fractions until the mass of the planet grows to facilitate advection mediated crossing. A fraction of these crossing solids are accreted by the planetary embryo, as they follow gas streamlines onto the accreting planet.}
    \label{fig:pcfiltered}
\end{figure}

To explore the fraction of solids able to cross the planet, we extend the integration of 1000 particles starting at 7 au to 1 Myr. We note here that over 1 Myr timescales, our assumptions that the disk structure and planet mass remain constant likely break down. However, the results presented here still give us a good insight to the total crossing efficiency for dust grains. While the fraction of solids crossing depends on time, shown in Figure \ref{fig:crosscdf}, we give the final fraction of solids that cross the orbit of the planet in Figure \ref{fig:pcfiltered}. In each figure, we show the fraction particles simulated that are either accreted by the protoplanet or cross from the outer disk to the inner disk. We also show the fraction particles that leave the bounds of our simulation, either by diffusing outward to the maximum radius of the simulation or by diffusing above or below $\pi/2 \pm 0.15$ in polar angle, the upper and lower bounds of our simulation. For the 10 and 100 \textmu m particles where particles diffuse out of bounds in the no planet case, additional particles are integrated for each planet mass until 1000 total particles drift inwards of 5.2 au by 1 Myr.
Consistent with previous work, we find that while the planetary embryo is less about 30 $M_\oplus$, particles larger than 1 mm drift inward past the growing protoplanet, while smaller particles tend to stay in the outer nebula for longer times, with about 30\% eventually diffusing inward. For more massive embryos, the larger particles are entirely trapped at the pressure bump exterior to the planet due to the local pressure maximum. However, up to 30\% of particles smaller than 100 \textmu m filter from the outer disk to the inner disk by 1 Myr, slightly increasing the crossing efficiency of these small particles as the planet grows.

At more massive embryo masses, small particles cross at higher efficiency than than low mass embryos because larger embryo masses drive stronger advective flows around the embryo (Figure \ref{fig:fargo_results}). Small particles are well coupled to the gas and follow this advective flow past the embryo and across the gap. As such, the meridional flows onto the embedded planet increase the relative crossing rate of these small particles. Note that at the largest planetary embryo only the smallest 10 \textmu m sized dust crosses the orbit of the planet. This is due to the increasing depth and width of the gas gap surrounding the planet. As the gas density around the embryo decreases, particle stokes numbers increase, such that only the smallest dust remains well coupled with the gas to follow the gas flows through the gap. Larger dust decouples from the gas before crossing into the gap.

We also find that a fraction of solids pass within the planetary envelope, given in equation \ref{eq:r_env}, meaning they are accreted by the planet. While the planet is below the pebble isolation mass, approximately 1 to 10\% of millimeter and centimeter sized pebbles are accreted, dependent on grain size and embryo mass, consistent with previous theoretical and modeling work on pebble accretion \citep{ormel_catching_2018, liu_catching_2018, drazkowska_how_2021, barnett_thermal_2022}.
Above the pebble isolation mass, the accreted solids come from the population of small dust in the disk, as these solids follow the meridional flow away from the midplane and onto the planet at higher altitudes. 
These results are supported by the findings of \citet{szulagyi_meridional_2022} and \citet{maeda_delivery_2024}, which show the primary accretion of material to an embedded giant embryo occurs through accretion of small dust at the polar regions of the embryo. Thus a growing Jovian-like planet regulates what solids it can accrete as it grows.

\subsection{Outward Transport of Solids}\label{sec:bothways}

Thus far we have focused on inward transport through the disk and across the planet's orbit, but outward transport of dust can occur as well. To investigate the importance of such movement we simulated the trajectories of 1 cm and 10 \textmu m sized grains (the largest and smallest sizes considered here) that originate in the range of 3.0 to 4.2 au and 6.2 to 7.4 au, considering particles that begin both inside and outside of the planetary embryo. While this ensures no particles begin within 1 au of the embryo, and thus outside the planetary envelope, some particles do begin in the gap opened by the most massive embryo. Such dust could be generated by collisions of dust aggregates or planetesimals present at these locations. We also choose to focus only on embryo masses above the pebble isolation mass, as this is where we expect the accretion flow mediated transport of small dust to be most prevalent.

\begin{figure}
    \centering\includegraphics[width=\linewidth]{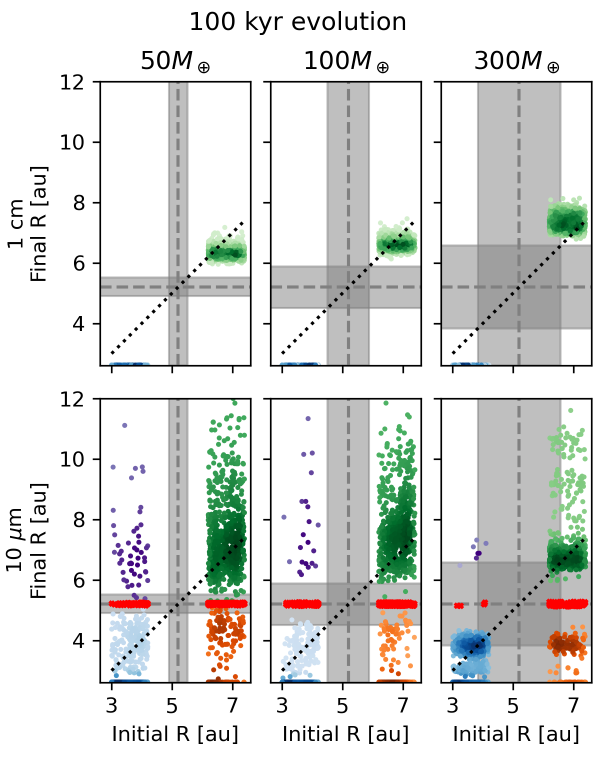}
    \caption{The starting and ending locations of solid particles that begin both inside and outside of the planet after 100 kyr of evolution. The planet location is shown by dashed grey lines, with the gap shown by the thick grey regions. A 1:1 line is shown by the black dotted line. Particles above this line moved outward in the nebula, while particles below this line moved inward. Points are colored according to the quadrant they are in, with darker colors indicated a higher density of points in that region of space. The upper right quadrant of each subfigure represents solids that start and end in the outer disk, while the bottom left are solids that start and end in the inner disk. The upper left and lower right quadrants show material that crosses the orbit of the planet, from the inside out or outside in respectively. Points in red show solids that are accreted by the planetary embryo.}
    \label{fig:startsandends}
\end{figure}

The initial radial locations of each particle and their positions after 100 kyr of evolution are shown in Figure \ref{fig:startsandends}. In the outer disk, 1 cm sized particles for each planet largely clump together at the pressure maxima where the gas flow is Keplerian. Particles of this size that start in the inner disk drift to the inner edge of the disk, and are removed from the simulation. The dynamics of these grains are largely dominated by gas drag induced drift. No cm sized grains cross the orbit of the planet for the embryo masses considered here.

The smaller, 10 \textmu m grains, shown in the bottom row of Figure \ref{fig:startsandends}, behave quite differently. For all embryo masses considered, the vast majority of small dust particles survive in the inner disk for the full 100 kyr due to their much longer drift timescales compared to the larger 1 cm size particles. In general, these smaller particles are dominated by diffusion rather than drift and tend to spread out through the disk rather than concentrate at a single location.
These solids are much more likely to cross the orbit of the growing planetary embryo in both directions, both moving inward from beyond the orbit of the planet as discussed above, but also outward from the inner disk to the outer disk. 
This affect can also be seen in \ref{fig:trajectories}, where 10 and 100 \textmu m sized grains can be seen to move between the inner and outer disk even in the presence of a 100 $M_\oplus$ embryo.
Thus, we find that some amount of small solids can be transported from the inner disk to the outer disk throughout the growth of the embedded embryo resulting in additional mixing of materials across the gap opened by the embryo.

As expected, the 1 cm particles tend to concentrate at the pressure bump where the solids move at the same speed as the gas resulting in no drag. These locations are at about 6.5, 7, and 8 au for the 50, 100 and 300 M$_\oplus$ cases respectively (see Figure \ref{fig:dvkep}). In the outer nebula, the smaller solids show a much wider distribution for the 50 and 100 M$_\oplus$ cases, although tend to concentrate at the gap edge for the 300 M$_\oplus$ case. This concentration of 10 \textmu m dust is due to the increased strength in gas flow due to gravitational interactions between gas and the massive embedded embryo. These dust concentration points are discussed in more detail in section \ref{sec:tricho}. Interestingly, these 10 \textmu m solids also tend to concentrate at the inner gap edge, at about 4 au.
At this location, the gas azimuthal velocity approaches Keplerian, although it is still slightly sub-Keplerian (see Fig \ref{fig:dvkep}). Additionally, at the gap edge, gas is flowing into the gap, radially outward at the inner gap edge. Because the gas is relatively depleted at this point in the disk, the small dust has a larger Stokes number than in the outer disk. As such, the dust do not diffuse as readily away from this region as would be expected in an unperturbed case. Combined, these effects result in a low radial velocity for the small dust, leading to an increase in concentration of small dust at the inner edge of the gap in addition to the outer edge. 

\subsection{Dust Concentration Points}\label{sec:tricho}

In the limit of many particles over long timescales, the residence times of the particles mirror the spatial density of particles at a given location. These regions of high dust concentration (long residence times) may be important growth regions, as the resulting elevated dust-to-gas ratios may lead to more frequent collisions \citep{birnstiel_simple_2012} and/or the initiation of the streaming instability \citep{youdin_streaming_2005, johansen_rapid_2007}.

The average residence time of a particle between radial location $R_j$ and $R_{j+1}$, where $R_j$ is the inner bin edge, is given by

\begin{equation}
    \tau_{res}(R_j) = \frac{1}{N_{part}}\sum_{i=0}^{N_{part}} \Delta t_{i} \delta(r_{i}),
\end{equation}
where $N_{part}$ is the number of particles, $\Delta t_i$ is the timestep of the integration, $r_i$ is the particle location at the start of the timestep, and

\begin{equation}
    \delta(r_i) = 
    \begin{cases}
        1, & \text{if $R_j < r_i < R_{j+1}$} \\
        0, & \text{else}
    \end{cases}.
\end{equation}

This can also be expanded to 2 or 3D residence times in Cartesian or spherical coordinates. The residence time in a given cell in the RZ plane is given by

\begin{equation}
    \tau_{res}(R_j,Z_k) = \frac{1}{N_{part}}\sum_{i=0}^{N_{part}} \Delta t_{i} \delta(r_{i}) \delta(z_{i}),
\end{equation}
where $\delta(z_{i})$ is defined similarly to $\delta(r_i)$ for a $z$-coordinate with bin edges at $Z_k$ and $Z_{k+1}$. These residence times are then normalized to the total integration time to give the fractional residence time.

\begin{figure}
    \centering
    \includegraphics[width=\linewidth]{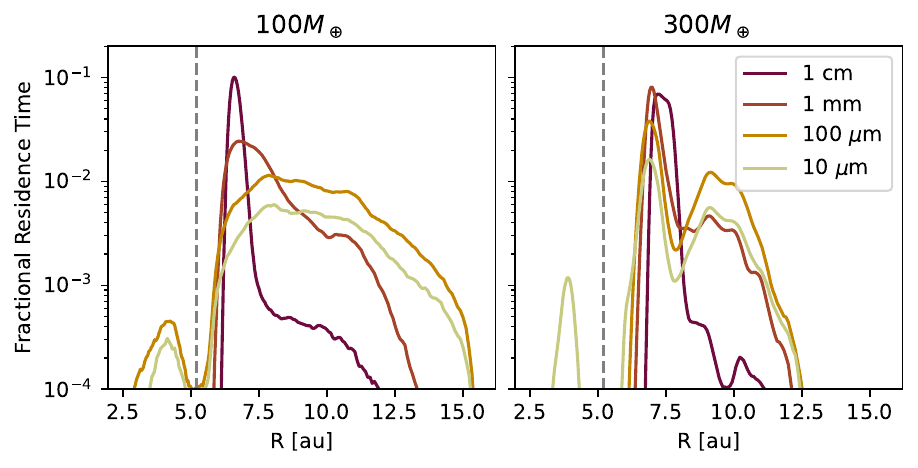}
    \caption{Average residence time for different sizes of solids as a function of radial location in the disk. Residence times are shown for disks containing planetary embryos of mass 100 $M_\oplus$ (left) and 300 $M_\oplus$ (right). For particles smaller than 1 mm, residence times show a two peaked distribution in the outer disk when $M=300\ M_\oplus$, but a smooth, single peaked distribution when $M=100\ M_\oplus$. The location of the planet at 5.2 au is indicated by the vertical dashed line in both plots. Thus, solids may form two distinct rings outside of the planet when the embedded Jovian embryo is massive enough.}
    \label{fig:restimes}
\end{figure}

We use the same bins for the residence times as are used for the FARGO3D mesh.  The radial residence times for the different sizes of particles considered for planetary embryo masses of 100 and 300 $M_\oplus$ are shown in Figure \ref{fig:restimes}. We use the same population of particles from section \ref{sec:filtering}, that is, solids initialized between 7 to 12 au over 100 kyr.

In both cases, 1 cm size particles tend to pile up in a narrow region just outside of the gap opened by the embryo, while the solids less than 1 mm in size are more spread out through the outer disk. While the smaller solids have a single peak distribution for the planetary embryo of 100 $M_\oplus$ around 7 au, the 300 $M_\oplus$ embryo creates a bimodal distribution of the small solids in the outer disk. This can be seen in the right panel of Figure \ref{fig:restimes}, with a sharp peak in residence time at about 7 au for all particles, (although the smaller particles tend to have their longest residence times closer to the planet). The solids smaller than 1 cm then also have a secondary broad peak centered at 10 au.
The massive planetary embryo creates global substructures throughout the disk, which can create concentric rings and gaps, as has been inferred for disks around other stars \citep{fedele_alma_2018, perez_dust_2019}. As such, a single giant planet may create multiple dust pileups in a single disk, explaining the observation of multiple ring and gap systems in other protoplanetary disks.

As shown in Figure \ref{fig:dvkep}, for embryo masses $30 M_\oplus \geq M_\text{pl} < 100 M_\oplus$, the azimuthal gas velocity is exactly Keplerian in only one location, corresponding to the pressure bump outside of the gap opened by the planet. But when $M_\text{pl} \gtrsim 200 M_\oplus$, there are multiple locations where $\delta v_\text{Kep} = 0$, and two locations where particles may concentrate. For a roughly Jupiter mass embryo of $300 M_\oplus$, these locations are at about 7.5 au, and 10 au.

We note that in the higher turbulence case shown here ($\alpha = 10^{-3}$), the outer pressure bump is not sufficient to trap the largest centimeter sized dust. However, smaller dust is more efficiently trapped in this outer pressure bump. 
This is because the larger dust is more vertically settled, and the dynamics are driven primarily by drift through the midplane of the disk \citep[see also][]{dullemond_disk_2018}. Thus, given the turbulence of the disk, a small diffusive ``kick'' can free the larger pebbles from the region where the gas velocity is Keplerian, and they rapidly drift inward. The smaller dust, on the other hand, is more vertically distributed and concentrates where the gas radial velocity is low or slightly outwards (offsetting inward drift). The small dust is much more sensitive to the radial gas advection not just along the midplane, but also up to a few scale heights in the disk.

\subsection{Dependence on Disk Viscosity}\label{sec:alpha4_results}

\begin{figure}
    \centering
    \includegraphics[width=\linewidth]{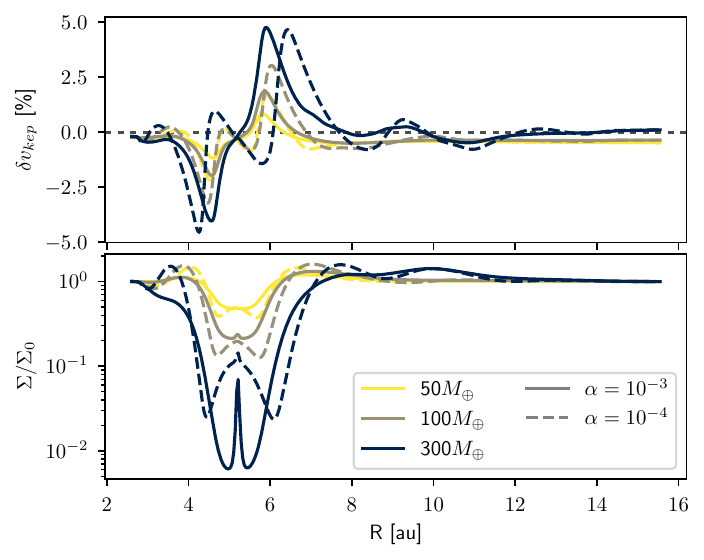}
    \caption{Comparison of azimuthally averaged disk structure for $\alpha = 10^{-3}$ (solid lines) and $10^{-4}$ (dashed lines). In general, gaps in disks with lower $\alpha$ values are both deeper and wider. In both cases, variations in the density remain in the outer disk, leading to multiple areas in the disks where azimuthal velocity is Keplerian. We note that while the region around the planet in the lower viscosity case always has lower density compared to the higher viscosity case, The 300 $M_\oplus$, case does not fully clear all of the gas from the gap in the timescales considered. The gas present in the gap is primarily opposite the planet (that is, $\phi=\pm \pi$). As particles crossing the orbit of the planet generally do so at the location of the planet, this gas present in the gap does not affect the crossing of particles.}
    \label{fig:a4_disk_structure}
\end{figure}

Although the level of turbulence in disks is uncertain, observations of PPDs have shown that turbulence can generally be modelled using an $\alpha$-disk approximation with $\alpha \approx 10^{-3} - 10^{-4}$ \citep[e.g.][]{doi_estimate_2021,flaherty_measuring_2020, flaherty_evidence_2024, paneque-carreno_high_2024}.
Here, we also explore a lower viscosity disk (i.e. lower turbulence) with $\alpha = 10^{-4}$. The lower viscosity affects the particle trajectories in two major ways. First, planets open deeper and wider gaps in the disk, as has been described in analytic and numerical studies \citep[e.g.][]{duffell_simple_2015, kanagawa_formation_2015, kanagawa_mass_2016}. The resulting azimuthally averaged gas structure of these $\alpha = 10^{-4}$ disks are shown compared to the fiducial $\alpha = 10^{-3}$ disks in Figure \ref{fig:a4_disk_structure}. Second, the diffusivity of the particles is also decreased, as the particle diffusivity is proportional to the gas viscosity (Equation \eqref{eq:part_diff}).

\begin{figure}
    \centering
    \includegraphics[width=\linewidth]{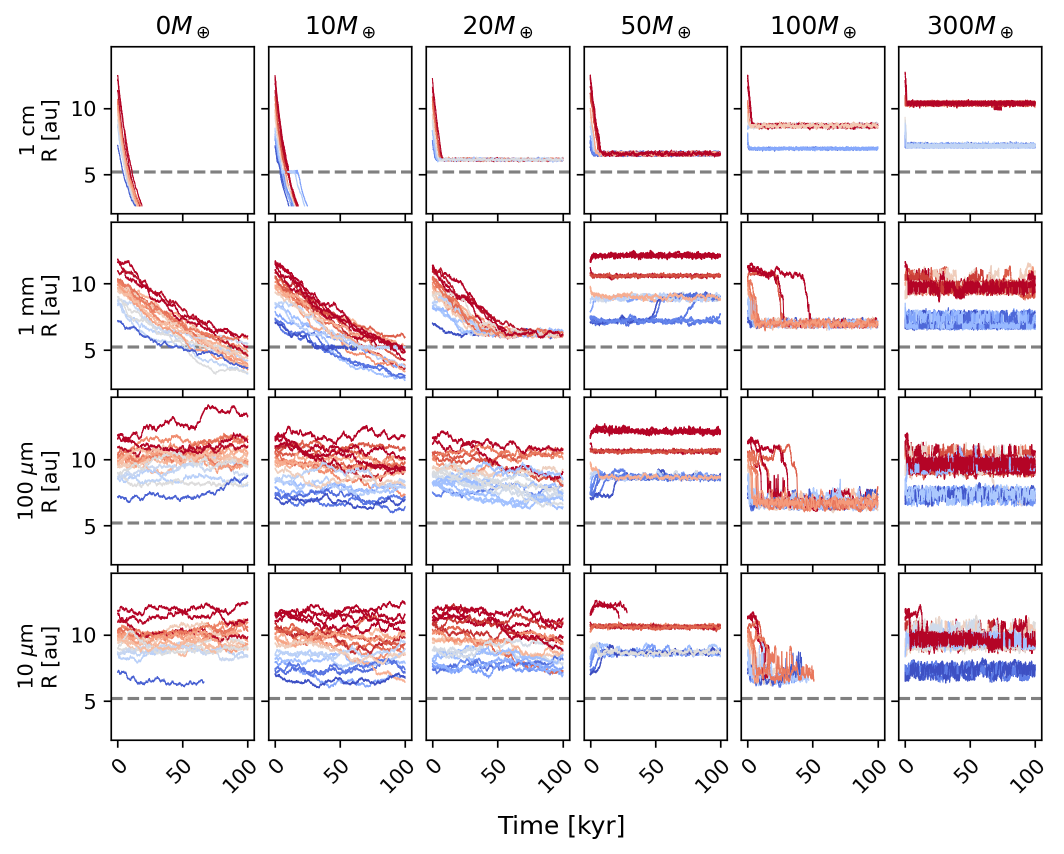}
    \caption{Some example trajectories of particles in a disk with an alpha of $10^{-4}$, as in figure \ref{fig:trajectories}. When compared with trajectories of solids in the \logalpha{3} disks, these solids are much more concentrated radially, and tend to group in multiple radial locations in the outer disk. Additionally, filtering from the outer disk to inner disk is much rarer, as particles become decoupled from the gas advection due to lower gas densities. We note that in some cases, all particles reach the edges of the simulation before the 100 kyr integration time. In these cases of low viscosity, this is due to large scale coherent flows in the disk, eventually bringing all particles above 3 scale heights, the upper-limit of our simulation.}
    \label{fig:a4_trajectories}
\end{figure}

The resulting trajectories of several particle integrations using $\alpha=10^{-4}$ are shown in Figure \ref{fig:a4_trajectories}. Here, similar to as in Section \ref{sec:filtering}, we focus on solids in the outer disk.
As can be seen, the lowered viscosity results in much less solid material crossing the orbit of the embedded embryo.
This finding is consistent with previous 2D models, which have shown the fraction of material crossing the gap opened by the planet decreases as the gas viscosity decreases \citep{weber_characterizing_2018, haugbolle_probing_2019}.
This is because the gas density at the edge of the gap is much lower than the $\alpha=10^{-3}$ case with a steeper density gradient, and so the grains decouple from the gas and no longer follow the gas advection across the gap. These solids are also unable to diffuse closer to the planet, where the gas advection and gravitational interactions may pull the solids towards the inner disk, as in the \logalpha{3} cases.  

\begin{figure}
    \centering
    \includegraphics[width=\linewidth]{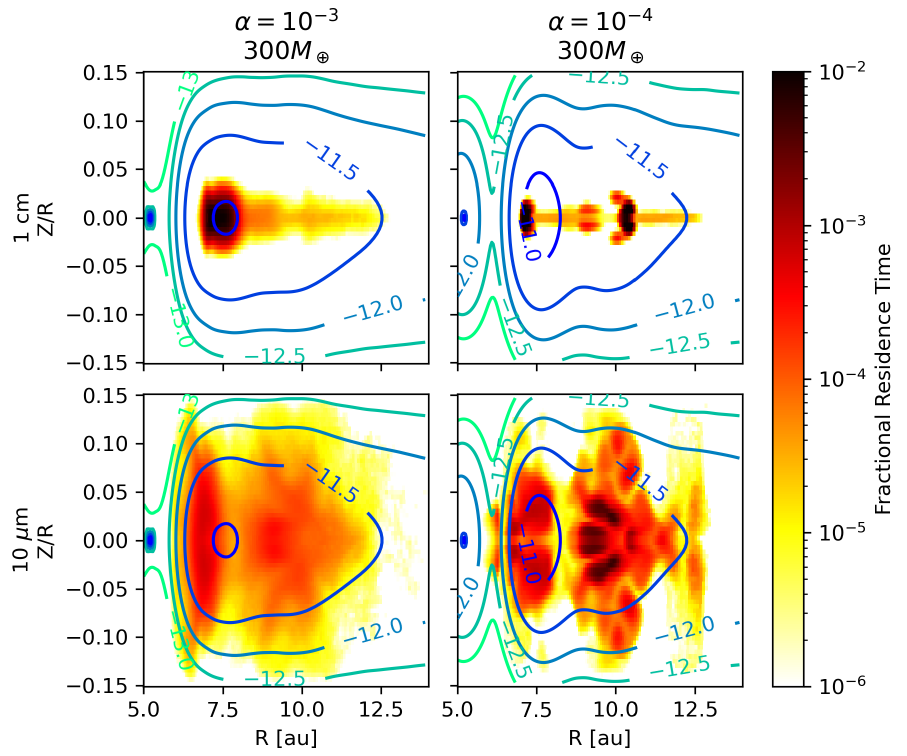}
    \caption{Azimuthally averaged residence times for the largest (top) and smallest (bottom) solids considered in high viscosity (left) and low viscosity (right) disks containing a 300 $M_\oplus$ planet. The contour lines show the surrounding gas density and are placed logarithmically, with contour labels indicating the $\log_10$ of the gas density. In the high viscosity case, the solids diffuse more vertically and tend to concentrate near the gap. In the high viscosity case, solids much more closely follow the gas advection, and are concentrated in much smaller areas. The small solids can spend significant amounts of time away from the midplane.}
    \label{fig:rz_restimes}
\end{figure}

Another feature of the lower viscosity cases is that the solids tend to concentrate at multiple locations at lower embryo masses when compared to the higher $\alpha=10^{-3}$ case. The lower viscosity leads to less damping of density and velocity perturbations in the outer disk, resulting in more pronounced substructure, as seen in the disk's azimuthal velocity and density shown in Figure \ref{fig:a4_disk_structure}. As a result, there are several regions of the disk that are at or near Keplerian velocity, leading to multiple pileups of all solids, including the drift-dominated, 1 cm size solids. Small dust that is well-coupled to the gas can also concentrate away from the midplane, as seen in Figure \ref{fig:rz_restimes}, leading to concentrations of the smaller 10 and 100 \textmu m dust. In lower viscosity disks, 
the embedded embryo can create large scale gas advection throughout the disk, with
gas moving away from the midplane and through the upper layers of the disk. For sufficiently small solids, they can become well entrained with these flows and concentrate at the center of these 3D gas circulation patterns, as shown by the regions of longer residence time in the lower right plot in figure \ref{fig:rz_restimes}.
Additionally, due to the lower particle diffusivity, the solids cannot easily diffuse away from these regions, and thus remain much more concentrated in space over the length of the particle integration.

\section{Discussion}\label{sec:discussion}

Given our results, we now explore how the growth of a giant embedded planet may affect the distribution of small dust grains and pebbles, including CAIs and other inclusions not yet incorporated into parent bodies, throughout the surrounding disk. In particular, we place these results in the context of Jupiter in the solar nebula and how the movement of solids around the embedded Jovian embryo can be reconciled with the meteoritic record as well as implications for other disks.

Jupiter is typically thought to serve as the barrier of NC and CC reservoirs in the disk, with NC interior to the Jovian embryo and CC exterior \citep[e.g.][]{kruijer_age_2017, kruijer_great_2020}. 
However, we have shown here that solids are able to mix between the inner and outer reservoirs, depending on the dust grain sizes and the mass of the growing embryo.
Not only can small solids can drift inwards past the embedded embryo, consistent with previous results, but we have also shown that there can be outward mixing through the nebula throughout the growth of Jupiter.
We explore how the results of the Monte Carlo technique presented here may relate to the mixing of isotopic reservoirs in the disk and the benefits of the Monte Carlo method presented here over alternative fluid-based simulations.

\subsection{Growth to the Pebble Isolation Mass}

We first consider the earliest stages of Jupiter’s growth, when it was below the pebble isolation mass.
In this early stage of growth, the embryo provides very little barrier to inward drift, with solids of all sizes crossing from the outer to the inner disk. As the embryo grows more massive, the transport of grains from the outer disk inward is slowed, with larger centimeter and millimeter sized grains completely trapped in the outer disk, while smaller sub-millimeter grains are able to cross the orbit of the planet even after the embryo has reached it's full Jupiter mass. The question arises, then, how Jupiter might act as a barrier between the inner NC and outer CC reservoirs throughout this time.
This is only possible if 1) dust growth is limited such that drift-dominated pebbles do not grow and 2) gas viscosity is low, preventing turbulent mixing of small grains (See Figures \ref{fig:trajectories}, \ref{fig:crosscdf}, \ref{fig:pcfiltered},  and \ref{fig:a4_trajectories}).  Given the ubiquity of millimeter-sized particles observed in other protoplanetary disks \citep[e.g.][]{andrews_disk_2018}, the presence of CAIs in meteorites in this size range, and the need for Jupiter’s core to grow to tens of Earth masses, this scenario seems unlikely.

Considering the presence of drift-dominated pebbles in the outer disk, transport of outer disk material to the inner disk is problematic for a number of reasons. If pebble accretion rates are indeed 1-10\% for large pebbles, as shown here and found in previous work \citep{ormel_catching_2018,liu_catching_2018,drazkowska_how_2021}, this implies at least 100 $M_\oplus$ of solids must drift past the proto-Jovian core before reaching 10 $M_\oplus$. Not only would this present a significant source of pollution for the the inner isotopic reservoir, but this would also supply the inner disk with much more mass than expected given the mass of the inner terrestrial planets \citep{brasser_partitioning_2020}.

We note that there are other models for Jupiter’s formation beyond pebble accretion.  Formation via gravitational instability has been suggested, but points to a highly dynamic environment where large-scale mixing across the disk would occur \citep[e.g.][]{boss_effect_2017, boss_effect_2019}, again making the preservation of separate isotopic reservoirs a challenge.  Accretion of Jupiter’s core via planetesimal collisions could also occur, but the question remains as to how to prevent the inward drift of solids during the ongoing growth of planetesimals, which we know formed over the course of millions of years \citep[e.g.][]{connelly_absolute_2012}. 

Given the challenge of preventing the inward drift of such a large amount of solids before Jupiter reached its isolation mass, we must consider that some other feature in the disk may have been responsible for creating a pressure bump in the disk early in its history.
A number of mechanisms have been proposed to explain the ring structures observed in disks beyond planets. For example, snowlines
may facilitate the growth of Jupiter while separating inner and outer reservoirs \citep{brasser_partitioning_2020, izidoro_planetesimal_2021, morbidelli_contemporary_2021}. As shown in \citet{lichtenberg_bifurcation_2021}, snowlines may also act as the seed locations for the NC and CC planetesimals, without the need for a gap carved by Jupiter. 
The formation of such a bump early on in the disk lifetime may indeed aid in the formation of giant planets such as Jupiter \citep{lau_sequential_2024}.
Thus, the isotopic dichotomy in the Solar Nebula may have been preserved by another process, and the formation of Jupiter was not the cause of the dichotomy, but an additional consequence of some other process or processes. More work considering the dynamical evolution of solids in such disks affected by these processes is needed.

\subsection{Growth Beyond the Pebble Isolation Mass}

We can also consider what the continued growth of Jupiter after it reached its isolation mass would mean for the preservation of the observed isotopic dichotomy.
After the embryo reached the isolation mass, the inward drift of large pebbles was halted in the outer disk, leading to a pileup of millimeter and larger solids. 
If at this time most of the solid mass was contained in these drift-dominated pebbles, then leaking between reservoirs would be limited.
Exchange of small dust, however, is possible in cases of high disk viscosity and turbulence.
As shown in Figure \ref{fig:pcfiltered}, we find that upwards of 10\% of small dust may have filtered from the outer disk to inner disk during this phase of Jupiter's growth for the case of $\alpha = 10^{-3}$. Additional transport of material from the inner reservoir outward may continue as shown in Figure \ref{fig:startsandends} and discussed in Section \ref{sec:bothways}.

Some amount of mixing between the inner and outer reservoirs at later stages of planet formation may help to explain various observations from the meteoritic record.
Previous work has found evidence of NC-like material in CC meteorites \citep{schrader_outward_2020, schrader_prolonged_2022, van_kooten_hybrid_2021} and interpreted this as a record of outward transport across Jupiter's orbit.  Further, presence of smaller CAIs in NC meteorites compared to CC meteorites \citep{dunham_calciumaluminum-rich_2023} could be consistent with the leakage of only small particles from the outer disk to the inner disk. Such outcomes are dependent on the turbulence in the disk, and as such we cannot rule out alternative explanations, for example, CAI storage in early forming planetesimals \citep{ebert_early_2024}.

It is important to note that the later-crossing small dust is facilitated by the growing embryo, and, as a result, any dust that crossed the orbit of Jupiter at this later period did so by passing nearby the accreting planet. We call this the ``funneling effect,'' as solids passing between inner and outer reservoirs are funneled past the embedded embryo. In some cases, we found particles moving within only a few hundred Jupiter radii of the embryo, just outside the planetary envelope (see Figure \ref{fig:onetrajectory}). While the details of the circumplanetary disk (or circumplanetary envelope) are not resolved here, the temperatures are expected to be elevated by $\sim100$ to $1000$~K depending on the mass accretion rate of the embryo \citep[see, e.g.,][]{szulagyi_accretion_2014,szulagyi_circumplanetary_2016, lega_gas_2024, krapp_thermodynamic_2024, alarcon_thermal_2024}. As such, later crossing dust may have lost volatiles as a result of thermal processing from the embedded embryo core \citep{barnett_thermal_2022}. If volatile ices and organics are the carriers of the supernova-derived isotopes \citep[e.g.][]{liu_natural_2022}, then this could provide a means of changing the isotopic compositions of the transported solids. The chemical implications of small grains passing by giant planets in the outer disk has recently been explored in \citet{petrovic_material_2024}.

\subsection{Global Effects and the Potential Trichotomy}

In addition to the dichotomy between NC and CC meteorites, Ti and Fe isotopic measurements of CI chondrites and recently returned samples from the asteroid Ryugu have revealed a third potential isotopic reservoir in the solar nebula \citep{tachibana_pebbles_2022,hopp_ryugus_2022,yokoyama_samples_2023}. As such, an alternative model of meteorite formation may be a trichotomy model, requiring further separation of dust into at least three distinct regions in the disk.

In our models, we see that as Jupiter grew, the embedded planet can create multiple concentric rings in the disk beyond the pressure bump immediately outside of the planetary gap. In our simulations this secondary pileup occurs at about 10 au, though its exact location will depend on where Jupiter formed. This secondary pressure bump can either slow the inward drift, or halt it completely depending on the viscosity of the disk and size of the grains. Importantly, this bump and the one immediately outside the planet source solids from distinct regions in the disk, as in Figures \ref{fig:trajectories} and \ref{fig:a4_trajectories}. If there was a compositional gradient in the initial distribution of solids then these multiple dust reservoirs in the outer disk may remain compositionally distinct from one another, though mixing between these outer disk reservoirs may occur depending on disk viscosity and turbulence.
These rings and the limited exchange of material between them may offer an explanation for the observed tertiary isotopic reservoir observed in CI meteorites and Ryugu samples.

Interestingly, we see that where grains do move between these reservoirs, they are transported quickly, moving multiple au through the disk on timescales much shorter than expected by drift. This can be seen in Figure \ref{fig:a4_trajectories} for 1 mm sized grains in disks containing 50 and 100 $M_\oplus$ embryos. This rapid transport may have important implications for understanding the chemistry of these solids as they change environments within the disk. The effects of UV radiation on ice-mantle chemistry of grains has been explored recently in, e.g. \citet{bergner_ice_2021} and \citet{flores-rivera_uv-processing_2024} using similar techniques.

The dynamical separation that develops at more distant regions from the planet is most effective when the viscosity of the disk is low, as higher disk viscosities both broaden pressure bumps in the disk and enable greater diffusion of solids out of these pressure bumps. This is in contrast with the hypothesized late inward exchange of CAIs and outward transport of NC material described above, which relies on a higher disk viscosity. If this transport did occur, preservation of the trichotomy may require additional effects or substructures. The subsequent formation of Saturn and the ice giants may act to further separate the CI reservoir in the outer disk \citep{hopp_ryugus_2022, nesvorny_isotopic_2024}.

\subsection{Future Work and Caveats}

While we have mainly focused on the role of Jupiter in the solar nebula, these results are also applicable to PPDs more broadly. For example, the mass available for planet formation is dependent on the pebble flux through the disk \citep{lambrechts_formation_2019, bitsch_formation_2019, mulders_why_2021}. Thus the presence of an outer giant planet may limit the pebble flux through the disk, reducing the occurrence of super-Earth planets in the inner disk. However, observations of exoplanetary systems reveal that systems with Jupiter-like exoplanets often also contain inner super-Earth planets \citep{zhu_super_2018, bryan_excess_2019, bryan_friends_2024}, although this correlation may not be conclusive \citep{barbato_exploring_2018}. The late, accretion-mediated filtering of small solids past an outer giant planet may allow enough small dust to pass that can then grow to pebbles, providing feedstock that would otherwise be shut off.

Although this work assumes constant gas density and velocity structure, this may be expanded in the future to allow for time varying disk conditions. This will allow for the study of the effects of migration, disk dissipation, and multiple planets. 
For example, the early disk may have been a more turbulent decretion disk, as opposed to the assumed standard accretion disk. Such structure may have been important for the early outward transport of CAIs \citep{cuzzi_blowing_2003,ciesla_outward_2007,yang_effects_2012,liu_natural_2022}.

Migration, in particular, may affect the fraction of material able to cross the orbit of the planet in either direction. In the regions of the disk explored here, planets tend to migrate inward faster than the unperturbed radial velocity of the gas \citep{duffell_migration_2014, durmann_migration_2015}. If the planet migrates faster than the surrounding gas, then the relative velocity of the gas with respect to the migrating planet should be negative, that is, from the inside the planet's orbit to outside. This relative outward gas velocity would inhibit the inward transport of small particles across the planet gap, and increase the outward transport of solids \citep{morbidelli_situ_2023}. This may increase the number of solids observed to migrate outwards as reported in Section \ref{sec:bothways}, providing a mechanism for introducing NC like material to the outer CC reservoir. We note, however, that small particles tend to cross the orbit of the planet nearby the planet (Figure \ref{fig:onetrajectory}), where the gas flow is onto the planet and into the gap, as shown in Figure \ref{fig:fargo_results}, rather than the gas velocity being only inward. As such, even with migration, we would still expect to see some amount of material crossing the orbit of the planet in both directions. That said, the effects of including a migrating planet with the particle tracking techniques presented here should be explored further in future work to address these questions on the filtering and accretion percentages. Migration would also affect the location in the disk where Jupiter is expected to form. For example, if Jupiter formed much further out in the disk than its current location \citep[i.e. beyond the N$_2$ snowline; ][]{oberg_jupiters_2019, bosman_jupiter_2019} then gas densities would be lower than explored here, affecting the meridional flows around the planet and the size of solids that remain well coupled to the gas.

Grain growth and fragmentation may have important implications for the transport of solids though the disk, as has been explored in, e.g.,\citet{drazkowska_including_2019, stammler_leaky_2023}. For example, fragmentation near the planet may increase the density of small particles, providing a source of small dust which could migrate across the planet's orbit,
as has recently been shown in \citet{eriksson_particle_2024}. Here, we do not explore the origin of the small dust. The consequences of the size evolution of solids should be explored more in future work, but it seems that fragmentation of large solids would increase the rate of solid mass transport across the planet's orbit.

As with all Monte Carlo methods, the results presented here are limited due to the finite number of particles integrated. In all cases, we explore a representative sample of a large number of particles originating in a small region of the disk. While we are not able to explore, for example, expected dust densities throughout the disk, this method excels in connecting regions of the disk, that is, understanding how particles from one region of the disk mix with material from other regions of the disk. Using the particle tracking technique presented here, we can begin to quantify the degree of radial mixing throughout the disk, as is discussed in Sections \ref{sec:filtering}, \ref{sec:bothways}, and \ref{sec:tricho}.
Although integrating a larger number of particles would enable us to better probe the low probability events (for example, accretion onto the planet or outward crossing particles), it would be computationally prohibitive to increase the number of particles explored. In its current implementation, the computation time scales linearly with the number of particles. As such, an order of magnitude increase in the number of particles integrated would require an order of magnitude increase in computation time.

Similar to limits on the number of particles, the computational time also scales linearly with the time of integration. As such, our simulations here are limited to less than 1 Myr, likely less than the age of the disk. However, it is important to note that throughout its lifetime, the disk and embedded planet are both constantly evolving. As such, the ``snapshots'' of constant disk and embryo mass are approximations, and such disk conditions are not expected to exist over million year timescales as considered. The results shown here, however, are useful limiting cases to demonstrate how mixing around the embedded planet changes over time.

Despite these drawbacks of Monte Carlo simulations, we believe there are benefits and novel insights of the techniques presented here compared with typical fluid-based studies. For one, we are able to study the dust evolution for a wide range of models out to 1 Myr. For our case with the planet at 5.2 au, this is equivalent to just under 100,000 orbits of the embryo, much longer than typical for hydrodynamic simulations. This is particularly true for the high resolution 3D hydrodynamic models we use for the gas structure here. Additionally, we are able to connect regions of the disk, quantifying mixing and accretion of solids onto the embedded embryo. While fluid approaches can identify regions of the disk where solids concentrate, similar to the residence time calculations in Section \ref{sec:tricho}, we have shown here that solids in a given ``clump'' tend to be sourced from nearby in the disk (see, Figures \ref{fig:a4_trajectories} and \ref{fig:rz_restimes} and discussion in Section \ref{sec:alpha4_results}) and, when they do move between reservoirs, do so quickly. These results are new findings of this technique and are not apparent from fluid approaches. Using these techniques we can begin to understand the chemical environments and residence times in those environments these are particles are expected to experience.

\section{Conclusion}\label{sec:conclusion}

In this work, we have combined 3D hydrodynamic simulations of a disk containing an embedded planetary embryo with particle tracking techniques to model the transport and trajectories of solids.
We place the histories of these small solids within the context of the meteoritic record of the Solar System, to constrain the timing and environment of Jupiter's formation. 

We find that there was a significant evolution and change in the dynamical behavior of dust as Jupiter grew.  Early on, before reaching the pebble isolation mass, solids were able to drift and diffuse from the outer disk to the inner disk freely and from all azimuthal angles.  Given the large flux of solids expected in the disk at this time, preserving the NC-CC dichotomy across the orbit of the planet would be challenging.  Instead, if some process or feature was already present that prevented or limited the drift of solids, then this would make preservation of the dichotomy possible and also could facilitate Jupiter's growth.

Once Jupiter grew to and beyond the pebble isolation mass, gas-drag-induced drift became inefficient, and the only solids that were capable of crossing the planet's orbit were small dust particles which were funneled from the outer disk to the inner disk, or from the inner to outer disk, immediately around the growing planet.  Given the presence of NC materials in CC meteorites \citep{schrader_outward_2020, schrader_prolonged_2022, van_kooten_hybrid_2021}, and that the only objects in the NC meteorites that are genetically linked to those in the CC meteorites, CAIs, are smaller and less abundant than found in their outer disk counterparts \citep{dunham_calciumaluminum-rich_2023}, it has been suggested that such crossing did occur.   If true, then this implies that the disk was sufficiently turbulent or viscous around the orbit of the planet.  Further, due to the funneling effect, 
any crossing solids passed nearby a luminous protoplanet, leading to some level of thermal processing of those solids.

Jupiter's growth would also impact the dynamics of dust particles at more distant regions of the solar nebula.  At sufficient mass, it could create an additional pressure bump in the outer disk, which would collect solids and source them from a different region than the bump immediately exterior to the planet's orbit.  This may further lead to compositional zoning of materials in the outer disk and may be recorded in the trichotomy suggested by analyses of CI meteorites and Ryugu samples.

Using the techniques presented here, we are able to identify individual grains that may cross the orbit of the embedded planet. Understanding the paths traveled by these solids will allow us to better understand the chemical processing expected to occur, as these solids move through drastically changing temperature and radiation environments throughout the disk. This technique also allows us to begin to quantify the extent of mixing between reservoirs in the disk and accretion onto the growing envelope and how these effects change over time as embedded planets grow. Thus, we can begin to study how planets sculpt their own environment during formation and affect the composition of the surrounding disk.

\section*{Acknowledgments}
This work was completed in part with resources provided by the University of Chicago’s Research Computing Center. The authors would like to thank Leonardo Krapp for advice regarding FARGO3D.
This material is based upon work supported by the National Aeronautics and Space Administration under Agreement No. 80NSSC21K0593 for the program “Alien Earths”. The results reported herein benefited from collaborations and/or information exchange within NASA’s Nexus for Exoplanet System Science (NExSS) research coordination network sponsored by NASA’s Science Mission Directorate.
EVC acknowledges support from NASA FINESST grant 80NSSC23K1380.
EMP thanks the Heising-Simons Foundation for support through the 51 Pegasi b Postdoctoral Fellowship.

\software{
\texttt{FARGO3D} \citep{benitez-llambay_fargo3d_2016, masset_fargo_2000}
\texttt{matplotlib} \citep{hunter_matplotlib_2007}
\texttt{numpy} \citep{harris_array_2020}
}

\bibliography{references}

\begin{thebibliography}{}
\expandafter\ifx\csname natexlab\endcsname\relax\def\natexlab#1{#1}\fi
\providecommand{\url}[1]{\href{#1}{#1}}
\providecommand{\dodoi}[1]{doi:~\href{http://doi.org/#1}{\nolinkurl{#1}}}
\providecommand{\doeprint}[1]{\href{http://ascl.net/#1}{\nolinkurl{http://ascl.net/#1}}}
\providecommand{\doarXiv}[1]{\href{https://arxiv.org/abs/#1}{\nolinkurl{https://arxiv.org/abs/#1}}}

\bibitem[{Alarcón \& Bergin(2024)}]{alarcon_thermal_2024}
Alarcón, F., \& Bergin, E.~A. 2024, The Astrophysical Journal, 967, 144, \dodoi{10.3847/1538-4357/ad3d57}

\bibitem[{Andrews {et~al.}(2018)Andrews, Huang, Pérez, Isella, Dullemond, Kurtovic, Guzmán, Carpenter, Wilner, Zhang, Zhu, Birnstiel, Bai, Benisty, Hughes, Öberg, \& Ricci}]{andrews_disk_2018}
Andrews, S.~M., Huang, J., Pérez, L.~M., {et~al.} 2018, The Astrophysical Journal, 869, L41, \dodoi{10.3847/2041-8213/aaf741}

\bibitem[{Armitage(2017)}]{armitage_lecture_2017}
Armitage, P.~J. 2017, arXiv:astro-ph/0701485.
\newblock \url{http://arxiv.org/abs/astro-ph/0701485}

\bibitem[{Bae \& Zhu(2018)}]{bae_planet-driven_2018}
Bae, J., \& Zhu, Z. 2018, The Astrophysical Journal, 859, 118, \dodoi{10.3847/1538-4357/aabf8c}

\bibitem[{Bae {et~al.}(2017)Bae, Zhu, \& Hartmann}]{bae_formation_2017}
Bae, J., Zhu, Z., \& Hartmann, L. 2017, The Astrophysical Journal, 850, 201, \dodoi{10.3847/1538-4357/aa9705}

\bibitem[{Barbato {et~al.}(2018)Barbato, Sozzetti, Desidera, Damasso, Bonomo, Giacobbe, Colombo, Lazzoni, Claudi, Gratton, LoCurto, Marzari, \& Mordasini}]{barbato_exploring_2018}
Barbato, D., Sozzetti, A., Desidera, S., {et~al.} 2018, Astronomy \& Astrophysics, 615, A175, \dodoi{10.1051/0004-6361/201832791}

\bibitem[{Barnett \& Ciesla(2022)}]{barnett_thermal_2022}
Barnett, M.~N., \& Ciesla, F.~J. 2022, The Astrophysical Journal, 925, 141, \dodoi{10.3847/1538-4357/ac4417}

\bibitem[{Benítez-Llambay \& Masset(2016)}]{benitez-llambay_fargo3d_2016}
Benítez-Llambay, P., \& Masset, F.~S. 2016, The Astrophysical Journal Supplement Series, 223, 11, \dodoi{10.3847/0067-0049/223/1/11}

\bibitem[{Bergner \& Ciesla(2021)}]{bergner_ice_2021}
Bergner, J.~B., \& Ciesla, F. 2021, The Astrophysical Journal, 919, 45, \dodoi{10.3847/1538-4357/ac0fd7}

\bibitem[{Binkert {et~al.}(2023)Binkert, Szulágyi, \& Birnstiel}]{binkert_three-dimensional_2023}
Binkert, F., Szulágyi, J., \& Birnstiel, T. 2023, Monthly Notices of the Royal Astronomical Society, 523, 55, \dodoi{10.1093/mnras/stad1405}

\bibitem[{Birnstiel {et~al.}(2012)Birnstiel, Klahr, \& Ercolano}]{birnstiel_simple_2012}
Birnstiel, T., Klahr, H., \& Ercolano, B. 2012, Astronomy \& Astrophysics, 539, A148, \dodoi{10.1051/0004-6361/201118136}

\bibitem[{Bitsch {et~al.}(2019)Bitsch, Izidoro, Johansen, Raymond, Morbidelli, Lambrechts, \& Jacobson}]{bitsch_formation_2019}
Bitsch, B., Izidoro, A., Johansen, A., {et~al.} 2019, Astronomy \& Astrophysics, 623, A88, \dodoi{10.1051/0004-6361/201834489}

\bibitem[{Bitsch {et~al.}(2018)Bitsch, Morbidelli, Johansen, Lega, Lambrechts, \& Crida}]{bitsch_pebble-isolation_2018}
Bitsch, B., Morbidelli, A., Johansen, A., {et~al.} 2018, Astronomy \& Astrophysics, 612, A30, \dodoi{10.1051/0004-6361/201731931}

\bibitem[{Bosman {et~al.}(2019)Bosman, Cridland, \& Miguel}]{bosman_jupiter_2019}
Bosman, A.~D., Cridland, A.~J., \& Miguel, Y. 2019, Astronomy \& Astrophysics, 632, L11, \dodoi{10.1051/0004-6361/201936827}

\bibitem[{Boss(2017)}]{boss_effect_2017}
Boss, A.~P. 2017, The Astrophysical Journal, 836, 53, \dodoi{10.3847/1538-4357/836/1/53}

\bibitem[{Boss(2019)}]{boss_effect_2019}
---. 2019, The Astrophysical Journal, 884, 56, \dodoi{10.3847/1538-4357/ab40a4}

\bibitem[{Brasser(2020)}]{brasser_partitioning_2020}
Brasser, R. 2020, Nature Astronomy, 4, 12

\bibitem[{Bryan {et~al.}(2019)Bryan, Knutson, Lee, Fulton, Batygin, Ngo, \& Meshkat}]{bryan_excess_2019}
Bryan, M.~L., Knutson, H.~A., Lee, E.~J., {et~al.} 2019, The Astronomical Journal, 157, 52, \dodoi{10.3847/1538-3881/aaf57f}

\bibitem[{Bryan \& Lee(2024)}]{bryan_friends_2024}
Bryan, M.~L., \& Lee, E.~J. 2024, The Astrophysical Journal Letters, 968, L25, \dodoi{10.3847/2041-8213/ad5013}

\bibitem[{Budde {et~al.}(2016)Budde, Burkhardt, Brennecka, Fischer-Gödde, Kruijer, \& Kleine}]{budde_molybdenum_2016}
Budde, G., Burkhardt, C., Brennecka, G.~A., {et~al.} 2016, Earth and Planetary Science Letters, 454, 293, \dodoi{10.1016/j.epsl.2016.09.020}

\bibitem[{Chambers(2017)}]{chambers_steamworlds_2017}
Chambers, J. 2017, The Astrophysical Journal, 849, 30, \dodoi{10.3847/1538-4357/aa91d0}

\bibitem[{Chambers(2021)}]{chambers_rapid_2021}
---. 2021, The Astrophysical Journal, 914, 102, \dodoi{10.3847/1538-4357/abfaa4}

\bibitem[{Ciesla(2007)}]{ciesla_outward_2007}
Ciesla, F.~J. 2007, Science, 318, 613, \dodoi{10.1126/science.1147273}

\bibitem[{Ciesla(2010)}]{ciesla_residence_2010}
---. 2010, The Astrophysical Journal, 723, 514, \dodoi{10.1088/0004-637X/723/1/514}

\bibitem[{Ciesla(2011)}]{ciesla_residence_2011}
---. 2011, The Astrophysical Journal, 740, 9, \dodoi{10.1088/0004-637X/740/1/9}

\bibitem[{Connelly {et~al.}(2012)Connelly, Bizzarro, Krot, Nordlund, Wielandt, \& Ivanova}]{connelly_absolute_2012}
Connelly, J.~N., Bizzarro, M., Krot, A.~N., {et~al.} 2012, Science, 338, 651, \dodoi{10.1126/science.1226919}

\bibitem[{Cuzzi {et~al.}(2003)Cuzzi, Davis, \& Dobrovolskis}]{cuzzi_blowing_2003}
Cuzzi, J.~N., Davis, S.~S., \& Dobrovolskis, A.~R. 2003, Icarus, 166, 385, \dodoi{10.1016/j.icarus.2003.08.016}

\bibitem[{Desch {et~al.}(2018)Desch, Kalyaan, \& Alexander}]{desch_effect_2018}
Desch, S.~J., Kalyaan, A., \& Alexander, C. M.~O. 2018, The Astrophysical Journal Supplement Series, 238, 11, \dodoi{10.3847/1538-4365/aad95f}

\bibitem[{Doi(2021)}]{doi_estimate_2021}
Doi, K. 2021, The Astrophysical Journal

\bibitem[{Drążkowska {et~al.}(2019)Drążkowska, Li, Birnstiel, Stammler, \& Li}]{drazkowska_including_2019}
Drążkowska, J., Li, S., Birnstiel, T., Stammler, S.~M., \& Li, H. 2019, The Astrophysical Journal, 885, 91, \dodoi{10.3847/1538-4357/ab46b7}

\bibitem[{Drążkowska {et~al.}(2021)Drążkowska, Stammler, \& Birnstiel}]{drazkowska_how_2021}
Drążkowska, J., Stammler, S.~M., \& Birnstiel, T. 2021, Astronomy \& Astrophysics, 647, A15, \dodoi{10.1051/0004-6361/202039925}

\bibitem[{Duffell(2015)}]{duffell_simple_2015}
Duffell, P.~C. 2015, The Astrophysical Journal, 807, L11, \dodoi{10.1088/2041-8205/807/1/L11}

\bibitem[{Duffell(2020)}]{duffell_empirically_2020}
---. 2020, The Astrophysical Journal, 889, 16, \dodoi{10.3847/1538-4357/ab5b0f}

\bibitem[{Duffell {et~al.}(2014)Duffell, Haiman, MacFadyen, D'Orazio, \& Farris}]{duffell_migration_2014}
Duffell, P.~C., Haiman, Z., MacFadyen, A.~I., D'Orazio, D.~J., \& Farris, B.~D. 2014, The Astrophysical Journal, 792, L10, \dodoi{10.1088/2041-8205/792/1/L10}

\bibitem[{Dullemond {et~al.}(2018)Dullemond, Birnstiel, Huang, Kurtovic, Andrews, Guzmán, Pérez, Isella, Zhu, Benisty, Wilner, Bai, Carpenter, Zhang, \& Ricci}]{dullemond_disk_2018}
Dullemond, C.~P., Birnstiel, T., Huang, J., {et~al.} 2018, The Astrophysical Journal Letters, 869, L46, \dodoi{10.3847/2041-8213/aaf742}

\bibitem[{Dunham {et~al.}(2023)Dunham, Sheikh, Opara, Matsuda, Liu, \& McKeegan}]{dunham_calciumaluminum-rich_2023}
Dunham, E.~T., Sheikh, A., Opara, D., {et~al.} 2023, Meteoritics \& Planetary Science, 58, 643, \dodoi{10.1111/maps.13975}

\bibitem[{Dürmann \& Kley(2015)}]{durmann_migration_2015}
Dürmann, C., \& Kley, W. 2015, Astronomy \& Astrophysics, 574, A52, \dodoi{10.1051/0004-6361/201424837}

\bibitem[{Ebert {et~al.}(2024)Ebert, Nagashima, Krot, Wakita, Barrat, \& Bischoff}]{ebert_early_2024}
Ebert, S., Nagashima, K., Krot, A.~N., {et~al.} 2024, Earth and Planetary Science Letters, 646, 119010, \dodoi{10.1016/j.epsl.2024.119010}

\bibitem[{Ebert {et~al.}(2018)Ebert, Render, Brennecka, Burkhardt, Bischoff, Gerber, \& Kleine}]{ebert_ti_2018}
Ebert, S., Render, J., Brennecka, G.~A., {et~al.} 2018, Earth and Planetary Science Letters, 498, 257, \dodoi{10.1016/j.epsl.2018.06.040}

\bibitem[{Eriksson {et~al.}(2024)Eriksson, Yang, \& Armitage}]{eriksson_particle_2024}
Eriksson, L. E.~J., Yang, C.-C., \& Armitage, P.~J. 2024, Monthly Notices of the Royal Astronomical Society: Letters, slae110, \dodoi{10.1093/mnrasl/slae110}

\bibitem[{Fedele {et~al.}(2018)Fedele, Tazzari, Booth, Testi, Clarke, Pascucci, Kospal, Semenov, Bruderer, Henning, \& Teague}]{fedele_alma_2018}
Fedele, D., Tazzari, M., Booth, R., {et~al.} 2018, Astronomy \& Astrophysics, 610, A24, \dodoi{10.1051/0004-6361/201731978}

\bibitem[{Flaherty {et~al.}(2020)Flaherty, Hughes, Simon, Qi, Bai, Bulatek, Andrews, Wilner, \& Kóspál}]{flaherty_measuring_2020}
Flaherty, K., Hughes, A.~M., Simon, J.~B., {et~al.} 2020, The Astrophysical Journal, 895, 109, \dodoi{10.3847/1538-4357/ab8cc5}

\bibitem[{Flaherty {et~al.}(2024)Flaherty, Hughes, Simon, Reina, Qi, Bai, Andrews, Wilner, \& Kóspál}]{flaherty_evidence_2024}
---. 2024, Monthly Notices of the Royal Astronomical Society, 532, 363, \dodoi{10.1093/mnras/stae1480}

\bibitem[{Flores-Rivera {et~al.}(2024)Flores-Rivera, Lambrechts, Gavino, Lorek, Flock, Johansen, \& Mignone}]{flores-rivera_uv-processing_2024}
Flores-Rivera, L., Lambrechts, M., Gavino, S., {et~al.} 2024, {UV}-processing of icy pebbles in the outer parts of {VSI}-turbulent disks,  arXiv, \dodoi{10.48550/arXiv.2412.01698}

\bibitem[{Fung \& Chiang(2016)}]{fung_gap_2016}
Fung, J., \& Chiang, E. 2016, The Astrophysical Journal, 832, 105, \dodoi{10.3847/0004-637X/832/2/105}

\bibitem[{Gerber {et~al.}(2017)Gerber, Burkhardt, Budde, Metzler, \& Kleine}]{gerber_mixing_2017}
Gerber, S., Burkhardt, C., Budde, G., Metzler, K., \& Kleine, T. 2017, The Astrophysical Journal, 841, L17, \dodoi{10.3847/2041-8213/aa72a2}

\bibitem[{Hammer {et~al.}(2017)Hammer, Kratter, \& Lin}]{hammer_slowly-growing_2017}
Hammer, M., Kratter, K.~M., \& Lin, M.-K. 2017, Monthly Notices of the Royal Astronomical Society, 466, 3533, \dodoi{10.1093/mnras/stw3000}

\bibitem[{Harris {et~al.}(2020)Harris, Millman, Walt, Gommers, Virtanen, Cournapeau, Wieser, Taylor, Berg, Smith, Kern, Picus, Hoyer, Kerkwijk, Brett, Haldane, Río, Wiebe, Peterson, Gérard-Marchant, Sheppard, Reddy, Weckesser, Abbasi, Gohlke, \& Oliphant}]{harris_array_2020}
Harris, C.~R., Millman, K.~J., Walt, S. J. v.~d., {et~al.} 2020, Nature, 585, 357, \dodoi{10.1038/s41586-020-2649-2}

\bibitem[{Haugbølle {et~al.}(2019)Haugbølle, Weber, Wielandt, Benítez-Llambay, Bizzarro, Gressel, \& Pessah}]{haugbolle_probing_2019}
Haugbølle, T., Weber, P., Wielandt, D.~P., {et~al.} 2019, The Astronomical Journal, 158, 55, \dodoi{10.3847/1538-3881/ab1591}

\bibitem[{Helled {et~al.}(2022)Helled, Stevenson, Lunine, Bolton, Nettelmann, Atreya, Guillot, Militzer, Miguel, \& Hubbard}]{helled_revelations_2022}
Helled, R., Stevenson, D.~J., Lunine, J.~I., {et~al.} 2022, Icarus, 378, 114937, \dodoi{10.1016/j.icarus.2022.114937}

\bibitem[{Hopp {et~al.}(2022)Hopp, Dauphas, Abe, Aléon, O’D.~Alexander, Amari, Amelin, Bajo, Bizzarro, Bouvier, Carlson, Chaussidon, Choi, Davis, Di~Rocco, Fujiya, Fukai, Gautam, Haba, Hibiya, Hidaka, Homma, Hoppe, Huss, Ichida, Iizuka, Ireland, Ishikawa, Ito, Itoh, Kawasaki, Kita, Kitajima, Kleine, Komatani, Krot, Liu, Masuda, McKeegan, Morita, Motomura, Moynier, Nakai, Nagashima, Nesvorný, Nguyen, Nittler, Onose, Pack, Park, Piani, Qin, Russell, Sakamoto, Schönbächler, Tafla, Tang, Terada, Terada, Usui, Wada, Wadhwa, Walker, Yamashita, Yin, Yokoyama, Yoneda, Young, Yui, Zhang, Nakamura, Naraoka, Noguchi, Okazaki, Sakamoto, Yabuta, Abe, Miyazaki, Nakato, Nishimura, Okada, Yada, Yogata, Nakazawa, Saiki, Tanaka, Terui, Tsuda, Watanabe, Yoshikawa, Tachibana, \& Yurimoto}]{hopp_ryugus_2022}
Hopp, T., Dauphas, N., Abe, Y., {et~al.} 2022, Science Advances, 8, eadd8141, \dodoi{10.1126/sciadv.add8141}

\bibitem[{Hunter(2007)}]{hunter_matplotlib_2007}
Hunter, J.~D. 2007, Computing in Science \& Engineering, 9, 90, \dodoi{10.1109/MCSE.2007.55}

\bibitem[{Izidoro {et~al.}(2021)Izidoro, Dasgupta, Raymond, Deienno, Bitsch, \& Isella}]{izidoro_planetesimal_2021}
Izidoro, A., Dasgupta, R., Raymond, S.~N., {et~al.} 2021, Nature Astronomy, \dodoi{10.1038/s41550-021-01557-z}

\bibitem[{Izquierdo {et~al.}(2022)Izquierdo, Facchini, Rosotti, Van~Dishoeck, \& Testi}]{izquierdo_new_2022}
Izquierdo, A.~F., Facchini, S., Rosotti, G.~P., Van~Dishoeck, E.~F., \& Testi, L. 2022, The Astrophysical Journal, 928, 2, \dodoi{10.3847/1538-4357/ac474d}

\bibitem[{Johansen {et~al.}(2007)Johansen, Oishi, Low, Klahr, Henning, \& Youdin}]{johansen_rapid_2007}
Johansen, A., Oishi, J.~S., Low, M.-M.~M., {et~al.} 2007, Nature, 448, 1022, \dodoi{10.1038/nature06086}

\bibitem[{Kanagawa {et~al.}(2016)Kanagawa, Muto, Tanaka, Tanigawa, Takeuchi, Tsukagoshi, \& Momose}]{kanagawa_mass_2016}
Kanagawa, K.~D., Muto, T., Tanaka, H., {et~al.} 2016, Publications of the Astronomical Society of Japan, 68

\bibitem[{Kanagawa {et~al.}(2015)Kanagawa, Tanaka, Muto, Tanigawa, \& Takeuchi}]{kanagawa_formation_2015}
Kanagawa, K.~D., Tanaka, H., Muto, T., Tanigawa, T., \& Takeuchi, T. 2015, Monthly Notices of the Royal Astronomical Society, 448, 994, \dodoi{10.1093/mnras/stv025}

\bibitem[{Krapp {et~al.}(2022)Krapp, Kratter, \& Youdin}]{krapp_3d_2022}
Krapp, L., Kratter, K.~M., \& Youdin, A.~N. 2022, The Astrophysical Journal, 928, 156, \dodoi{10.3847/1538-4357/ac5899}

\bibitem[{Krapp {et~al.}(2024)Krapp, Kratter, Youdin, Benítez-Llambay, Masset, \& Armitage}]{krapp_thermodynamic_2024}
Krapp, L., Kratter, K.~M., Youdin, A.~N., {et~al.} 2024, The Astrophysical Journal, 973, 153, \dodoi{10.3847/1538-4357/ad644a}

\bibitem[{Kruijer {et~al.}(2017)Kruijer, Burkhardt, Budde, \& Kleine}]{kruijer_age_2017}
Kruijer, T.~S., Burkhardt, C., Budde, G., \& Kleine, T. 2017, Proceedings of the National Academy of Sciences, 114, 6712, \dodoi{10.1073/pnas.1704461114}

\bibitem[{Kruijer {et~al.}(2020)Kruijer, Kleine, \& Borg}]{kruijer_great_2020}
Kruijer, T.~S., Kleine, T., \& Borg, L.~E. 2020, Nature Astronomy, 4, 32, \dodoi{10.1038/s41550-019-0959-9}

\bibitem[{Lambrechts {et~al.}(2014)Lambrechts, Johansen, \& Morbidelli}]{lambrechts_separating_2014}
Lambrechts, M., Johansen, A., \& Morbidelli, A. 2014, Astronomy \& Astrophysics, 572, A35, \dodoi{10.1051/0004-6361/201423814}

\bibitem[{Lambrechts {et~al.}(2019)Lambrechts, Morbidelli, Jacobson, Johansen, Bitsch, Izidoro, \& Raymond}]{lambrechts_formation_2019}
Lambrechts, M., Morbidelli, A., Jacobson, S.~A., {et~al.} 2019, Astronomy \& Astrophysics, 627, A83, \dodoi{10.1051/0004-6361/201834229}

\bibitem[{Lau {et~al.}(2024)Lau, Birnstiel, Drążkowska, \& Stammler}]{lau_sequential_2024}
Lau, T. C.~H., Birnstiel, T., Drążkowska, J., \& Stammler, S.~M. 2024, Astronomy \& Astrophysics, 688, A22, \dodoi{10.1051/0004-6361/202450464}

\bibitem[{Lega {et~al.}(2024)Lega, Benisty, Cridland, Morbidelli, Schulik, \& Lambrechts}]{lega_gas_2024}
Lega, E., Benisty, M., Cridland, A., {et~al.} 2024, Gas dynamics around a {Jupiter}-mass planet {I}. {Influence} of protoplanetary disk properties,  arXiv.
\newblock \url{http://arxiv.org/abs/2408.12233}

\bibitem[{Lichtenberg {et~al.}(2021)Lichtenberg, Drążkowska, Schönbächler, Golabek, \& Hands}]{lichtenberg_bifurcation_2021}
Lichtenberg, T., Drążkowska, J., Schönbächler, M., Golabek, G., \& Hands, T. 2021, Science, 371, 365

\bibitem[{Liu {et~al.}(2022)Liu, Johansen, Lambrechts, Bizzarro, \& Haugbølle}]{liu_natural_2022}
Liu, B., Johansen, A., Lambrechts, M., Bizzarro, M., \& Haugbølle, T. 2022, Science Advances, 8, eabm3045, \dodoi{10.1126/sciadv.abm3045}

\bibitem[{Liu \& Ormel(2018)}]{liu_catching_2018}
Liu, B., \& Ormel, C.~W. 2018, Astronomy \& Astrophysics, 615, A138, \dodoi{10.1051/0004-6361/201732307}

\bibitem[{Maeda {et~al.}(2024)Maeda, Ohtsuki, Suetsugu, Shibaike, Tanigawa, \& Machida}]{maeda_delivery_2024}
Maeda, N., Ohtsuki, K., Suetsugu, R., {et~al.} 2024, The Astrophysical Journal, 968, 62, \dodoi{10.3847/1538-4357/ad4035}

\bibitem[{Masset(2000)}]{masset_fargo_2000}
Masset, F. 2000, Astronomy and Astrophysics Supplement Series, 141, 165, \dodoi{10.1051/aas:2000116}

\bibitem[{Morbidelli {et~al.}(2021)Morbidelli, Baillié, Batygin, Charnoz, Guillot, Rubie, \& Kleine}]{morbidelli_contemporary_2021}
Morbidelli, A., Baillié, K., Batygin, K., {et~al.} 2021, Nature Astronomy, 6, 72, \dodoi{10.1038/s41550-021-01517-7}

\bibitem[{Morbidelli {et~al.}(2023)Morbidelli, Batygin, \& Lega}]{morbidelli_situ_2023}
Morbidelli, A., Batygin, K., \& Lega, E. 2023, Astronomy \& Astrophysics, 675, A75, \dodoi{10.1051/0004-6361/202346868}

\bibitem[{Morbidelli {et~al.}(2014)Morbidelli, Szulágyi, Crida, Lega, Bitsch, Tanigawa, \& Kanagawa}]{morbidelli_meridional_2014}
Morbidelli, A., Szulágyi, J., Crida, A., {et~al.} 2014, Icarus, 232, 266, \dodoi{10.1016/j.icarus.2014.01.010}

\bibitem[{Mulders {et~al.}(2021)Mulders, Drążkowska, Van Der~Marel, Ciesla, \& Pascucci}]{mulders_why_2021}
Mulders, G.~D., Drążkowska, J., Van Der~Marel, N., Ciesla, F.~J., \& Pascucci, I. 2021, The Astrophysical Journal Letters, 920, L1, \dodoi{10.3847/2041-8213/ac2947}

\bibitem[{Nesvorný {et~al.}(2024)Nesvorný, Dauphas, Vokrouhlický, Deienno, \& Hopp}]{nesvorny_isotopic_2024}
Nesvorný, D., Dauphas, N., Vokrouhlický, D., Deienno, R., \& Hopp, T. 2024, Earth and Planetary Science Letters, 626, 118521, \dodoi{10.1016/j.epsl.2023.118521}

\bibitem[{Ormel \& Liu(2018)}]{ormel_catching_2018}
Ormel, C.~W., \& Liu, B. 2018, Astronomy \& Astrophysics, 615, A178, \dodoi{10.1051/0004-6361/201732562}

\bibitem[{Paneque-Carreño {et~al.}(2024)Paneque-Carreño, Izquierdo, Teague, Miotello, Bergin, Loomis, \& Van~Dishoeck}]{paneque-carreno_high_2024}
Paneque-Carreño, T., Izquierdo, A.~F., Teague, R., {et~al.} 2024, Astronomy \& Astrophysics, 684, A174, \dodoi{10.1051/0004-6361/202347757}

\bibitem[{Petrovic {et~al.}(2024)Petrovic, Booth, \& Clarke}]{petrovic_material_2024}
Petrovic, H.~J., Booth, R.~A., \& Clarke, C.~J. 2024, Material {Transport} in {Protoplanetary} {Discs} with {Massive} {Embedded} {Planets},  arXiv.
\newblock \url{http://arxiv.org/abs/2409.16245}

\bibitem[{Pinilla {et~al.}(2012)Pinilla, Benisty, \& Birnstiel}]{pinilla_ring_2012}
Pinilla, P., Benisty, M., \& Birnstiel, T. 2012, Astronomy \& Astrophysics, 545, A81, \dodoi{10.1051/0004-6361/201219315}

\bibitem[{Pinte {et~al.}(2018)Pinte, Price, Ménard, Duchêne, Dent, Hill, Gregorio-Monsalvo, Hales, \& Mentiplay}]{pinte_kinematic_2018}
Pinte, C., Price, D.~J., Ménard, F., {et~al.} 2018, The Astrophysical Journal, 860, L13, \dodoi{10.3847/2041-8213/aac6dc}

\bibitem[{Pinte {et~al.}(2019)Pinte, van~der Plas, Ménard, Price, Christiaens, Hill, Mentiplay, Ginski, Choquet, Boehler, Duchêne, Perez, \& Casassus}]{pinte_kinematic_2019}
Pinte, C., van~der Plas, G., Ménard, F., {et~al.} 2019, Nature Astronomy, 3, 1109, \dodoi{10.1038/s41550-019-0852-6}

\bibitem[{Pérez {et~al.}(2019)Pérez, Casassus, Baruteau, Dong, Hales, \& Cieza}]{perez_dust_2019}
Pérez, S., Casassus, S., Baruteau, C., {et~al.} 2019, The Astronomical Journal, 158, 15, \dodoi{10.3847/1538-3881/ab1f88}

\bibitem[{Schrader \& Davidson(2022)}]{schrader_prolonged_2022}
Schrader, D.~L., \& Davidson, J. 2022, Earth and Planetary Science Letters, 589, 117552, \dodoi{10.1016/j.epsl.2022.117552}

\bibitem[{Schrader {et~al.}(2020)Schrader, Nagashima, Davidson, McCoy, Ogliore, \& Fu}]{schrader_outward_2020}
Schrader, D.~L., Nagashima, K., Davidson, J., {et~al.} 2020, Geochimica et Cosmochimica Acta, 282, 133, \dodoi{10.1016/j.gca.2020.05.014}

\bibitem[{Shakura \& Sunyaev(1973)}]{shakura_black_1973}
Shakura, N.~I., \& Sunyaev, R.~A. 1973, Astronomy and Astrophysics, 24, 337.
\newblock \url{https://ui.adsabs.harvard.edu/abs/1973A&A....24..337S}

\bibitem[{Stammler {et~al.}(2023)Stammler, Lichtenberg, Drążkowska, \& Birnstiel}]{stammler_leaky_2023}
Stammler, S.~M., Lichtenberg, T., Drążkowska, J., \& Birnstiel, T. 2023, Astronomy \& Astrophysics, 670, L5, \dodoi{10.1051/0004-6361/202245512}

\bibitem[{Szulágyi {et~al.}(2022)Szulágyi, Binkert, \& Surville}]{szulagyi_meridional_2022}
Szulágyi, J., Binkert, F., \& Surville, C. 2022, The Astrophysical Journal, 924, 1, \dodoi{10.3847/1538-4357/ac32d1}

\bibitem[{Szulágyi {et~al.}(2016)Szulágyi, Masset, Lega, Crida, Morbidelli, \& Guillot}]{szulagyi_circumplanetary_2016}
Szulágyi, J., Masset, F., Lega, E., {et~al.} 2016, Monthly Notices of the Royal Astronomical Society, 460, 2853, \dodoi{10.1093/mnras/stw1160}

\bibitem[{Szulágyi {et~al.}(2014)Szulágyi, Morbidelli, Crida, \& Masset}]{szulagyi_accretion_2014}
Szulágyi, J., Morbidelli, A., Crida, A., \& Masset, F. 2014, The Astrophysical Journal, 782, 65, \dodoi{10.1088/0004-637X/782/2/65}

\bibitem[{Tachibana {et~al.}(2022)Tachibana, Sawada, Okazaki, Takano, Sakamoto, Miura, Okamoto, Yano, Yamanouchi, Michel, Zhang, Schwartz, Thuillet, Yurimoto, Nakamura, Noguchi, Yabuta, Naraoka, Tsuchiyama, Imae, Kurosawa, Nakamura, Ogawa, Sugita, Morota, Honda, Kameda, Tatsumi, Cho, Yoshioka, Yokota, Hayakawa, Matsuoka, Sakatani, Yamada, Kouyama, Suzuki, Honda, Yoshimitsu, Kubota, Demura, Yada, Nishimura, Yogata, Nakato, Yoshitake, Suzuki, Furuya, Hatakeda, Miyazaki, Kumagai, Okada, Abe, Usui, Ireland, Fujimoto, Yamada, Arakawa, Connolly, Fujii, Hasegawa, Hirata, Hirata, Hirose, Hosoda, Iijima, Ikeda, Ishiguro, Ishihara, Iwata, Kikuchi, Kitazato, Lauretta, Libourel, Marty, Matsumoto, Michikami, Mimasu, Miura, Mori, Nakamura-Messenger, Namiki, Nguyen, Nittler, Noda, Noguchi, Ogawa, Ono, Ozaki, Senshu, Shimada, Shimaki, Shirai, Soldini, Takahashi, Takei, Takeuchi, Tsukizaki, Wada, Yamamoto, Yoshikawa, Yumoto, Zolensky, Nakazawa, Terui, Tanaka, Saiki, Yoshikawa, Watanabe, \& Tsuda}]{tachibana_pebbles_2022}
Tachibana, S., Sawada, H., Okazaki, R., {et~al.} 2022, Science, 375, 1011, \dodoi{10.1126/science.abj8624}

\bibitem[{Takeuchi \& Lin(2002)}]{takeuchi_radial_2002}
Takeuchi, T., \& Lin, D. N.~C. 2002, The Astrophysical Journal, 581, 1344, \dodoi{10.1086/344437}

\bibitem[{Tanigawa {et~al.}(2014)Tanigawa, Maruta, \& Machida}]{tanigawa_accretion_2014}
Tanigawa, T., Maruta, A., \& Machida, M.~N. 2014, The Astrophysical Journal, 784, 109, \dodoi{10.1088/0004-637X/784/2/109}

\bibitem[{Teague {et~al.}(2019)Teague, Bae, \& Bergin}]{teague_meridional_2019}
Teague, R., Bae, J., \& Bergin, E.~A. 2019, Nature, 574, 378, \dodoi{10.1038/s41586-019-1642-0}

\bibitem[{Teague {et~al.}(2018)Teague, Bae, Bergin, Birnstiel, \& Foreman-Mackey}]{teague_kinematical_2018}
Teague, R., Bae, J., Bergin, E.~A., Birnstiel, T., \& Foreman-Mackey, D. 2018, The Astrophysical Journal, 860, L12, \dodoi{10.3847/2041-8213/aac6d7}

\bibitem[{Van~Kooten {et~al.}(2021)Van~Kooten, Schiller, Moynier, Johansen, Haugbølle, \& Bizzarro}]{van_kooten_hybrid_2021}
Van~Kooten, E., Schiller, M., Moynier, F., {et~al.} 2021, The Astrophysical Journal, 910, 70, \dodoi{10.3847/1538-4357/abd9c8}

\bibitem[{Warren(2011)}]{warren_stable-isotopic_2011}
Warren, P.~H. 2011, Earth and Planetary Science Letters, 311, 93, \dodoi{10.1016/j.epsl.2011.08.047}

\bibitem[{Weber {et~al.}(2018)Weber, Benítez-Llambay, Gressel, Krapp, \& Pessah}]{weber_characterizing_2018}
Weber, P., Benítez-Llambay, P., Gressel, O., Krapp, L., \& Pessah, M.~E. 2018, The Astrophysical Journal, 854, 153, \dodoi{10.3847/1538-4357/aaab63}

\bibitem[{Weidenschilling(1977)}]{weidenschilling_aerodynamics_1977}
Weidenschilling, S.~J. 1977, Monthly Notices of the Royal Astronomical Society, 180, 57, \dodoi{10.1093/mnras/180.2.57}

\bibitem[{Yang \& Ciesla(2012)}]{yang_effects_2012}
Yang, L., \& Ciesla, F.~J. 2012, Meteoritics \& Planetary Science, 47, 99, \dodoi{10.1111/j.1945-5100.2011.01315.x}

\bibitem[{Yokoyama {et~al.}(2023)Yokoyama, Nagashima, Nakai, Young, Abe, Aléon, Alexander, Amari, Amelin, Bajo, Bizzarro, Bouvier, Carlson, Chaussidon, Choi, Dauphas, Davis, Rocco, Fujiya, Fukai, Gautam, Haba, Hibiya, Hidaka, Homma, Hoppe, Huss, Ichida, Iizuka, Ireland, Ishikawa, Ito, Itoh, Kawasaki, Kita, Kitajima, Kleine, Komatani, Krot, Liu, Masuda, McKeegan, Morita, Motomura, Moynier, Nguyen, Nittler, Onose, Pack, Park, Piani, Qin, Russell, Sakamoto, Schönbächler, Tafla, Tang, Terada, Terada, Usui, Wada, Wadhwa, Walker, Yamashita, Yin, Yoneda, Yui, Zhang, Connolly, Lauretta, Nakamura, Naraoka, Noguchi, Okazaki, Sakamoto, Yabuta, Abe, Arakawa, Fujii, Hayakawa, Hirata, Hirata, Honda, Honda, Hosoda, Iijima, Ikeda, Ishiguro, Ishihara, Iwata, Kawahara, Kikuchi, Kitazato, Matsumoto, Matsuoka, Michikami, Mimasu, Miura, Morota, Nakazawa, Namiki, Noda, Noguchi, Ogawa, Ogawa, Okada, Okamoto, Ono, Ozaki, Saiki, Sakatani, Sawada, Senshu, Shimaki, Shirai, Sugita, Takei, Takeuchi, Tanaka, Tatsumi, Terui, Tsuda,
  Tsukizaki, Wada, Watanabe, Yamada, Yamada, Yamamoto, Yano, Yokota, Yoshihara, Yoshikawa, Yoshikawa, Furuya, Hatakeda, Hayashi, Hitomi, Kumagai, Miyazaki, Nakato, Nishimura, Soejima, Suzuki, Yada, Yamamoto, Yogata, Yoshitake, Tachibana, \& Yurimoto}]{yokoyama_samples_2023}
Yokoyama, T., Nagashima, K., Nakai, I., {et~al.} 2023, Science, 379, eabn7850, \dodoi{10.1126/science.abn7850}

\bibitem[{Youdin \& Goodman(2005)}]{youdin_streaming_2005}
Youdin, A.~N., \& Goodman, J. 2005, The Astrophysical Journal, 620, 459, \dodoi{10.1086/426895}

\bibitem[{Youdin \& Lithwick(2007)}]{youdin_particle_2007}
Youdin, A.~N., \& Lithwick, Y. 2007, Icarus, 192, 588, \dodoi{10.1016/j.icarus.2007.07.012}

\bibitem[{Zhu \& Wu(2018)}]{zhu_super_2018}
Zhu, W., \& Wu, Y. 2018, The Astronomical Journal, 156, 92, \dodoi{10.3847/1538-3881/aad22a}

\bibitem[{Zhu {et~al.}(2014)Zhu, Stone, Rafikov, \& Bai}]{zhu_particle_2014}
Zhu, Z., Stone, J.~M., Rafikov, R.~R., \& Bai, X.-n. 2014, The Astrophysical Journal, 785, 122, \dodoi{10.1088/0004-637X/785/2/122}

\bibitem[{Öberg \& Wordsworth(2019)}]{oberg_jupiters_2019}
Öberg, K.~I., \& Wordsworth, R. 2019, The Astronomical Journal, 158, 194, \dodoi{10.3847/1538-3881/ab46a8}

\bibitem[{Öberg {et~al.}(2021)Öberg, Guzmán, Walsh, Aikawa, Bergin, Law, Loomis, Alarcón, Andrews, Bae, Bergner, Boehler, Booth, Bosman, Calahan, Cataldi, Cleeves, Czekala, Furuya, Huang, Ilee, Kurtovic, Le~Gal, Liu, Long, Ménard, Nomura, Pérez, Qi, Schwarz, Sierra, Teague, Tsukagoshi, Yamato, van~’t Hoff, Waggoner, Wilner, \& Zhang}]{oberg_molecules_2021}
Öberg, K.~I., Guzmán, V.~V., Walsh, C., {et~al.} 2021, The Astrophysical Journal Supplement Series, 257, 1, \dodoi{10.3847/1538-4365/ac1432}

\end{thebibliography}
\bibliographystyle{aasjournal}

\appendix

\section{HD Disk Structure}\label{sec:3d-structure}

The modeled disk has the surface density and scale height relation as given in section \ref{sec:method-hd}, with surface density power-law slope $b=-1$, aspect ratio $A = H/R_0 = 0.05$, and flaring index $\gamma=1/4$. Initially, the disk is axisymmetric (and thus independent of azimuthal angle $\phi$) and in hydrostatic equilibrium. The density of each cell at radius $r$ and polar angle $\theta$ is given by
\begin{equation}
    \rho(r,\theta) = \frac{\Sigma_0}{\sqrt{2\pi}}\frac{1}{R_0A}\left(\frac{r}{R_0}\right)^{-\xi}\sin(\theta)^{-\xi-\beta}\exp\left[(1-\sin(\theta)^{-2\gamma}\right)/2\gamma h^2]
\end{equation}
where
\begin{align}
    \xi &= b+1+\gamma \\
    \beta &= 1-2\gamma \\
    h &= \frac{H}{r} = A\left(\frac{r}{R_0}\right)^\gamma
\end{align}

The FARGO3D setup file can be provided upon request.

\end{document}